\documentclass[aps, prd, superscriptaddress, nofootinbib]{revtex4}

\usepackage[colorlinks=true, linkcolor=red, citecolor=blue]{hyperref}

\usepackage{graphicx}
\usepackage{amsmath,amssymb,amsfonts,mathtools}

\newcommand{\be}{\begin{equation}}
\newcommand{\ee}{\end{equation}}
\newcommand{\bea}{\begin{eqnarray}}
\newcommand{\eea}{\end{eqnarray}}
\newcommand{\kp}{$\kappa$-Poincar\'e }

\newcommand{\D}{\text{d}} %differential d
\newcommand{\arccosh}{\text{arccosh}}

\begin{document}

\title{Relativistic compatibility of the interacting $\kappa$-Poincar\'e model and implications for the relative locality framework}

\author{Giulia Gubitosi}

\affiliation{Departamento de F\'isica, Universidad de Burgos, E-09001 Burgos, Spain}

\author{Sjors Heefer}

\affiliation{Institute for Mathematics, Astrophysics and Particle Physics, Radboud University
Heyendaalseweg 135, 6525 AJ Nijmegen, The Netherlands}

\affiliation{Department of Mathematics and Computer Science, Eindhoven University of Technology, Eindhoven, The Netherlands}

\begin{abstract}
We investigate the relativistic properties under boost transformations of the $\kappa$-Poincar\'e model with multiple causally connected interactions, both at the level of its formulation in momentum space only and when it is endowed with a full phase space construction, provided by the relative locality framework. Previous studies focussing on the momentum space picture showed that in presence of just one interaction vertex the model is relativistic, provided that the boost parameter acting on each given particle receives a ``backreaction'' from the momenta of the other particles that participate in the interaction. Here we show that in presence of multiple causally-connected vertices the model is relativistic if the boost parameter acting on each given particle receives a backreaction from the total momentum of all the particles that are causally connected, even those that do not directly enter the vertex. The relative locality framework constructs spacetime by defining a set of dual coordinates to the momentum of each particle and interaction coordinates as Lagrange multipliers that enforce momentum conservation at interaction events. We show that the picture is Lorentz invariant if one uses an appropriate ``total boost'' to act on the particles' worldlines and on the interaction coordinates. The picture we develop also allows for a reinterpretation of the backreaction as the manifestation of the ``total boost'' action. Our findings provide the basis to consistently define distant relatively boosted observers in the relative locality framework.
\end{abstract}
\maketitle

\section{Introduction}

In the past few years models of Planck-scale deformed special relativity (DSR) \cite{AmelinoCamelia:2010pd} and their realization in terms of theories with Planck-scale curved momentum space \cite{KowalskiGlikman:2003we, KowalskiGlikman:2002ft} have been playing an increasingly relevant role in quantum gravity research.

On the one hand the theoretical stance of such models is now based on quite firm grounds: indications that the Planck-scale structure of spacetime implies a deformation of the geometry of momentum space emerge in research on noncommutative geometry \cite{Majid:1999tc, AmelinoCamelia:1999pm, KowalskiGlikman:2002jr}, loop quantum gravity \cite{Amelino-Camelia:2016gfx, Cianfrani:2016ogm} as well as $2+1$ dimensional quantum gravity \cite{Matschull:1997du, Meusburger:2003ta, Freidel:2005me}. 
On the other hand, the prospects for testing Planck-scale deviations from special relativity are now more concrete than ever:  analyses concerning  the time of flight of very high energy particles of astrophysical origin have  reached the required Planck-scale sensitivity, and   found regularities \cite{Amelino-Camelia:2015nqa,  Amelino-Camelia:2016ohi,  Xu:2016zxi,  Amelino-Camelia:2017zva,  Huang:2018ham, Xu:2018ien} that are compatible with the  sort of energy dependence of the velocity of massless particles that would be expected in  typical DSR  scenarios  (while the extremely strong constraints on deviations from standard physics in threshold reactions disfavour  the Lorentz breaking (LIV) scenario  \cite{Jacobson:2005bg, AmelinoCamelia:2008qg}).   

The recent deepening in the understanding of relativistic models with  curved momentum space geometry has led us to appreciate the highly nontrivial and counterintuitive implications of deformed  special relativity. Probably the most striking one is the necessity to abandon the standard concept of absolute locality: an interaction can only be established to be local  by  nearby observers, while distant observers might  see the interaction as nonlocal \cite{AmelinoCamelia:2010qv, AmelinoCamelia:2011bm}. This feature can be exposed once one develops a relativistically compatible spacetime picture alongside the momentum space construction. This was achieved with the relative locality framework, which takes momentum space as the base manifold and defines  spacetime as the cotangent space to a point on the  manifold \cite{AmelinoCamelia:2011bm,AmelinoCamelia:2011pe}, see Section \ref{sec:RL} for a brief review.

A much studied model for  Planck-scale-deformed momentum space geometry is the so-called  $\kappa$-Poincar\'e model, whose symmetries are compatible with those generated by the  $\kappa$-Poincar\'e Hopf algebra \cite{Lukierski:1991pn, Lukierski:1993wxa, Majid:1994cy}, and whose associated relative locality construction of spacetime was presented in \cite{Gubitosi:2013rna,AmelinoCamelia:2011nt}.
The popularity of this model can be ascribed to the fact that its Hopf algebra foundation provides a mathematically consistent framework to describe deformed relativistic symmetries and its very rich structure allows us to expose with clarity the sort of challenges that can arise when deforming Special Relativity. 
We will use this model as basis for our investigation, and start by reviewing the construction of the \kp momentum space model from the \kp Hopf algebra in Section \ref{sec:symmetries_kp_momentum_space}.

Because, as mentioned, the most promising phenomenology to date concerns propagation of free particles, most of the theoretical efforts have concentrated on  understanding the noninteracting version of the \kp model (and of DSR models in general). As will be reviewed in Subsection \ref{sub:kPfree}, the specific way in which relativistic invariance is realised by the \kp momentum space is well understood at the single particle level:  symmetry transformations leave the dispersion relation invariant. Relative locality effects can also be exposed very clearly at this level, by comparing how the simultaneous emission or detection of particles is seen by different observers \cite{AmelinoCamelia:2010qv,Amelino-Camelia:2013uya}. 

Introducing interactions generates additional complexity, because  relativistic consistency requires  to account for the interplay between deformed translation invariance (responsible for the deformation of the conservation rules of momenta) and deformed Lorentz invariance \cite{Carmona:2011wc}. From a Hopf-algebraic point of view,  this is due to the nontrivial co-algebraic sector of the Hopf algebra. The nontriviality of the translation sector reflects into the deformed composition rule of momenta, while the nontriviality of the boost sector produces a mixing between Lorentz and translation generators (see the beginning of Section \ref{sec:symmetries_kp_momentum_space}). First clues about the kinds of features that one should expect in interacting models were already uncovered  in  studies of systems with one interaction vertex, which focussed on the momentum space only: the transformation relating relatively boosted observers depends on the momentum of the particles involved in the vertex, so that the rapidity with which each particle  is boosted receives a ``backreaction'' from the momenta of the other particles participating  in the interaction  \cite{Gubitosi:2013rna}. We review this in Subsection \ref{sub:backreaction}. 

The interplay between deformed translations and boosts is expected to have especially virulent effects in presence of several interactions. In this case deformed  Lorentz transformations  need to be consistent with deformed translational invariance, manifested not only as a deformed conservation rule of momenta, but also as a deformed transformation law relating far away observers, each local  to one of the interaction events in order to make reliable inferences \cite{Amelino-Camelia:2014qaa}. In Subsection \ref{sub:kPmultivertex} we report our advancements on these issues. First, we show that  the commutation rules between translation and boost generators are such that the transformation linking distant relatively boosted observers involves a momentum-dependent translation. Moreover, we present the generalization of the relativistic picture developed in the single-interaction framework to the case where particles undergo several causally connected interactions. The conceptual difficulty resides in conciliating the previously-established notion that the rapidity parameter acting on a particle depends on the momenta of the other particles in the interaction vertex with the fact that a given particle can participate to more than one vertex, so that according to each vertex the particle should have a different rapidity.  We find that the solution to the paradox is to realize that in fact the rapidity is affected by \emph{all} the causally connected momenta: the rapidity of a particle taking part to a chain of interactions receives a backreaction from the momenta of all particles involved, not just those that are directly interacting with it (i.e. the transformation rule between two relatively boosted observers depends on the  momentum of all causally connected particles).
We also find that this prescription can only work for a certain class of interaction chains, and we argue that the requirement of relativistic invariance  lets us select the physically allowed interaction chains. These turn out to be the interaction chains that also preserve global momentum conservation, previously identified in \cite{Amelino-Camelia:2014qaa}.

Having established the relativistic compatibility of the interacting $\kappa$-Poincar\'e momentum space model in greater generality than before, we go on to discuss whether this model allows for a relativistically compatible spacetime picture within the relative locality framework.  This is an issue that was never addressed before, even in the single-interaction case.  As discussed in Section \ref{sec:RL}, in presence of several interacting particles the relative locality framework defines spacetime coordinates for each particle as the cotangent space to the point in momentum space representing the momentum of the particle. Moreover the framework  provides a prescription for stating that the particles are actually interacting, since it also defines ``interaction coordinates" as the Lagrange multipliers enforcing momentum conservation. This construction was demonstrated to be compatible with translational invariance \cite{AmelinoCamelia:2011nt}, but compatibility with boost invariance was never addressed. In Section \ref{sec:RLrelativistic} we  are able to complete the understanding of  interactions from the relative locality perspective, showing how Lorentz invariance is achieved in a nontrivial way. Indeed, we show that the relativistically compatible Lorentz transformation of a system of interacting particles is generated by the ``total boost'' generator, which is a nontrivial sum (provided by the structure of the underlying Hopf algebra) of the boost generators acting on single particles. This total boost generator governs the transformation of the spacetime coordinates of each particle, as well as that of the interaction coordinates. When more than one interaction vertex is present, the total boost accounts for all of the causally connected particles. Interestingly, we show that this prescription allows for a reinterpretation of the transformation rule of the particles' momenta, since the backreaction of momenta over rapidity can be seen as a manifestation of the action of the total boost.

We conclude the paper by showing how observers can be defined once one is able to build a network of causally connected events compatible with the relativistic symmetries of the model. In particular, we discuss the nontrivial relation between distant and relatively boosted observers and show how these observers assign  different amounts of non-locality to interactions belonging to a causally connected chain.

In this paper we work in  $1+1$ dimensions, since this is sufficient to expose the main results and conceptual innovations of our work. The generalization to the $3+1$ dimensional case is not expected to entail significant additional  difficulties. We also adopt units such that $c=\hbar=1$.

\section{Relativistic compatibility of the  $\kappa$-Poincar\'e momentum space model}\label{sec:symmetries_kp_momentum_space}

As mentioned in the introduction, the \kp momentum space model is based on the symmetries of the \kp Hopf algebra. In this Section we will review briefly how this momentum space is constructed and what are its main properties.  In the  bicrossproduct basis \cite{Majid:1994cy}  of the \kp algebra   the generators associated to spacetime translations, $P_0$, $P_1$, and boost, $N$, have the following commutation relations:
\be
[ P_0, P_1] = 0\,,\qquad [ N, P_0] =  P_1\,, \qquad [ N,  P_1] = \frac{\kappa}{2}\left(1-e^{-2 P_0/\kappa}\right) - \frac{1}{2\kappa} P_1^2\,, \label{eq:kp_algebra}
\ee
and coalgebra:
\be
\label{eq:kp_coalgebra}
\Delta( P_0) =  P_0\otimes 1 + 1 \otimes  P_0\,,\qquad \Delta( P_1) =  P_1 \otimes 1 + e^{- P_0/\kappa} \otimes  P_1\,, \qquad \Delta( N) =  N \otimes 1 + e^{- P_0/\kappa} \otimes  N\,.
\ee
The parameter $\kappa$, with dimensions of a momentum, governs the deformation with respect to the classical Poincar\'e algebra, which is recovered in the $\kappa^{-1}\rightarrow 0$ limit. Because of the connection to quantum gravity research, the parameter $\kappa$ is expected to be roughly of the order of the Planck scale $E_{p}\simeq 10^{28}$ eV.

Other relevant  structures of the Hopf algebra are the counit
\be
\epsilon( P_0) = \epsilon( P_1) = \epsilon( N) = 0\,, 
\ee
and the antipode,
\be
S( P_0) = - P_0\,,\qquad S( P_1) = - e^{ P_0/\kappa}  P_1 \,,\qquad S( N) = - e^{ P_0/\kappa}  N\,.
\ee
Finally, the Casimir element is
\be
C = 4 \kappa ^2 \sinh^2\left(\frac{P_0}{2\kappa}\right)-(P_1)^2 e^{P_0/\kappa }\,.
\ee

Because in the bicrossproduct basis the translation generators form a Hopf-subalgebra, they can be represented as an algebra of functions over momentum space \cite{KowalskiGlikman:2003we, Gubitosi:2013rna}, such that the  two translation generators correspond to the coordinate functions $p_0$ and $p_1$,
\be
\label{eq:Prepresentation}
P_0 = p_0,\qquad P_1 = p_1\,.
\ee
Then one can establish a correspondence between the structures of the Hopf sub-algebra of translations and the properties of the momentum space, thus providing a physical interpretation of the Hopf algebra mathematical construction. Specifically, a deformed composition law of momenta is read off from the Hopf algebra coproduct:
\be 
(p\oplus q)_\mu = (\Delta (P_\mu))(p,q) \qquad\Rightarrow\qquad \left\{ \begin{array}{lcl}
(p\oplus q)_0 &=& p_0 + q_0\,,\\
 (p\oplus q)_1 &=& p_1 + e^{-p_0/\kappa}q_1\,.
\end{array}
\right. \label{eq:kP_composition_law}
\ee
This is associative (because of the coassociativity of the coproduct) but noncommutative (because of the noncocommutativity of the coproduct). The internal structure of the Hopf algebra guarantees that the composition law makes momentum space into a group with unit element provided by the co-unit, 
\be 
\tilde 0_\mu=\epsilon(P_\mu)=0\,,
\ee
and inverse element provided by the antipode:
\be
(\ominus p)_\mu = (S(P_\mu))(p)\qquad \Rightarrow \qquad \left\{
\begin{array}{lcl}
(\ominus p)_0 &=& -p_0\,,\\
(\ominus p)_1 &=& -e^{p_0/\kappa}p_1\,.
\end{array}\right. 
\ee
Indeed, one can easily check that for any $p_\mu$
\be
p\oplus  \tilde 0 =  \tilde 0 \oplus p = p,\qquad p\oplus(\ominus p) = (\ominus p)\oplus p =  \tilde 0.
\ee

From the Hopf algebra structure  one can also infer the mass-shell condition, which is naturally identified with the Casimir, since this object is the invariant associated to the Hopf algebra. Upon representing the Casimir on momentum space one finds the dispersion relation:
\be
\label{eq:dispersion_relation}
m^2 = 4 \kappa ^2 \sinh^2\left(\frac{p_0}{2\kappa}\right)-(p_1)^2 e^{p_0/\kappa }.
\ee

The momentum space thus constructed has the geometry of (half of) a de Sitter manifold, with curvature given by the parameter $\kappa$ \cite{KowalskiGlikman:2003we, Gubitosi:2013rna, Amelino-Camelia:2013uya}. The link to the  phenomenological applications mentioned in the introduction is established when this momentum space model is taken to describe the kinematics and dynamics of classical particles, where ``classical" refers to a regime where purely quantum effects, such as worldline fuzziness, can be neglected.  

Since the \kp momentum space model just constructed is based on a Hopf algebra, it is reasonable to expect that it is relativistically consistent, in the sense that it allows to describe (systems of) particles in a way that preserves  relativistic invariance under the deformed  Poincar\'e transformations defined by this algebra. However  relativistic invariance  might be realized in  physically nontrivial ways. As we  discuss below, this is indeed the case when looking at particles undergoing an interaction. Because of the nontrivial co-algebraic sector of the \kp algebra, it turns out that the action of boosts on a given particle depends on the other particles interacting with it.  This Section is devoted to review known successes of the \kp model in describing a relativistic framework as far as free particles and systems with one interaction vertex are concerned. Adding more complexity to the picture, we also study systems of multiple causally connected interactions. In this case the challenge is to understand the interplay between the nontrivial Lorentz  transformations and the translations that connect the different interaction events.  
 
 \subsection{Free particle}\label{sub:kPfree}

The relativistic invariance of a  free  particle model is quite straightforward. In momentum space the only relevant object is the particle's dispersion relation, Eq. \eqref{eq:dispersion_relation}. Because it is derived from the Casimir of the algebra, we expect it to be trivially invariant, given that the Casimir by definition commutes with all the generators of the symmetry transformations. And indeed we can explicitly show that when boosting the particle's momentum via\footnote{As mentioned above we adopt a semiclassical approximation, so the action of generators on the momentum space functions is via Poisson brackets. The properties of the generators of the Hopf algebra are inherited by the Poisson brackets  with the convention that if  $[G, f(P_{\mu})]=h(P_{\mu})$, then $\{G,f(p_{\mu})\} = h(p_{\mu})$ for any generator $G$ of the Hopf algebra. The functions $f$, $h$ take as argument  the translation generators $P_{\mu}$ in the first case and the momentum space coordinates $p_{\mu}$ in the second case.}
\bea
p_0&\to&B^{\xi} \triangleright p_{0} \equiv p_0 + \xi\{N,p_0\} = p_0 + \xi p_1,\nonumber \\
p_1&\to& B^{\xi} \triangleright p_{1} \equiv p_1+ \xi\{N,p_1\} = p_1 + \xi \left[\frac{\kappa}{2}\left(1-e^{-2p_0/\kappa}\right) - \frac{1}{2\kappa}p_1^2\right], \label{eq:boosted_momenta}
\eea
in which $\xi$ is the rapidity parameter and we consider infinitesimal boost transformations, the dispersion relation is left unchanged, since
\be\label{eq:commutator_boost_with_casimir}
\{N,4 \kappa ^2 \sinh^2\left(\frac{p_0}{2\kappa}\right)-(p_1)^2 e^{p_0/\kappa } \}=0\,.
\ee
Because we are focussing on a momentum space model and no reference to spacetime is made at this level, invariance under translations $T_{a}$, where $a=\{a^{0},a^{1}\}$ is the translation parameter, is trivial since momenta are left unchanged by translations:
\be
p_{\nu}\rightarrow T_{a}\triangleright p_{\nu}\equiv p_{\nu}+a^{\mu}\left\{P_{\mu},p_{\nu}\right\} =p_{\nu}\,.\label{eq:translated_momenta}
\ee

\subsection{One interaction vertex}
\label{sub:backreaction}

When considering interactions matters become more involved. In fact, already when only one interaction vertex is present, one has to account for the interplay between translational invariance (which manifests itself in the deformed momentum composition law, Eq. \eqref{eq:kP_composition_law}) and invariance under Lorentz transformations.

\begin{figure}[h]
\begin{center}
\includegraphics[width=0.5\textwidth]{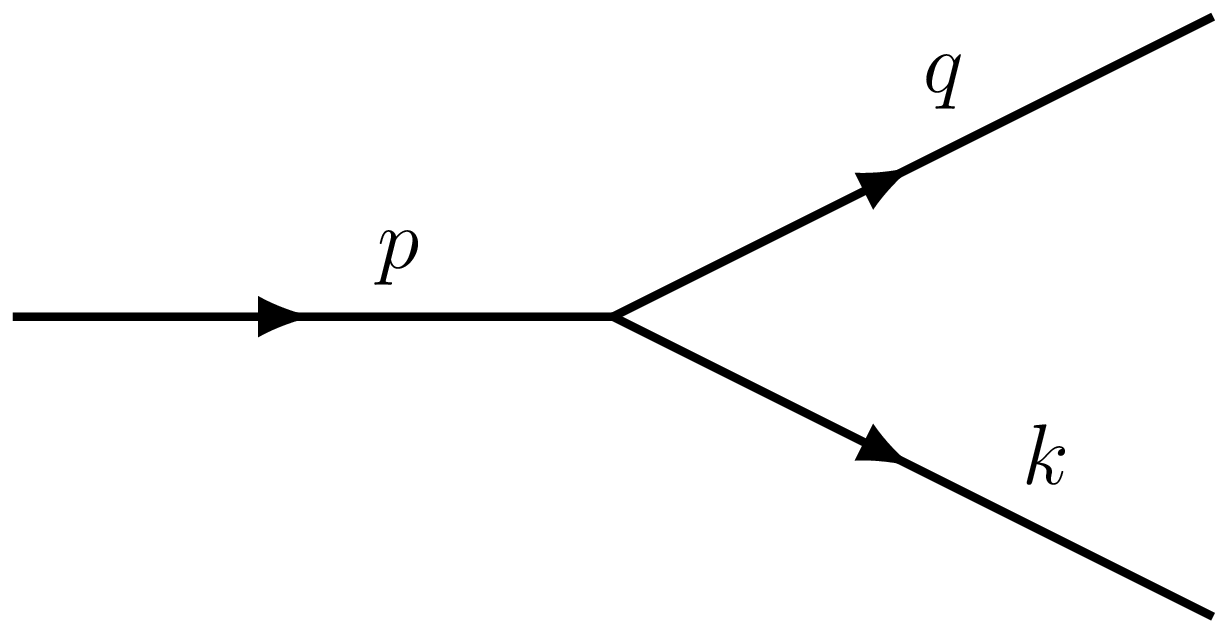}
\caption{Interaction diagram of one incoming particle with momentum $p$ and two outgoing particles with momenta $q$ and $k$. The lines in the diagram do not represent the actual worldlines of the particles, but simply indicate how the momentum is distributed among particles (in a similar spirit as in the Feynman diagrams, even though here particles are classical). Time evolution is read off going from left to right. The order of (noncommutative) summation of momenta is also encoded in the diagram, with the convention that momenta are summed from the top to the bottom. In this specific diagram one sums the momenta $q$ and $k$ as $q\oplus k$.}
\label{fig:p_to_q_k}
\end{center}
\end{figure}

We  show how relativistic invariance is achieved in this case via an explicit example: that of a particle with momentum $p$ which decays into two particles, with momentum $q$ and $k$ respectively. Such process is represented in Figure \ref{fig:p_to_q_k}. Note that because the addition law of momenta, Eq. \eqref{eq:kP_composition_law}, is noncommutative, one has to specify the order in which the momenta of the outgoing particles are composed. We  adopt the convention that the momenta of particles depicted at the top enter the composition law before those of particles depicted at the bottom. Making reference to Figure \ref{fig:p_to_q_k}, this means that the conservation law reads:
\be
p_0 = q_0 + k_0,\qquad p_1 = q_1 + e^{-q_0/\kappa}k_1.
\ee
It is now well understood \cite{Gubitosi:2013rna} (see also \cite{AmelinoCamelia:2011yi}), that such momentum conservation law is not covariant under the most straightforward way of boosting the interacting momenta, which is to boost each momentum with the same rapidity $\xi$:
\be
p\to B^{\xi} \triangleright p\,,\quad q\to B^{\xi} \triangleright q \,,\quad k\to B^{\xi} \triangleright k\,.
\ee
In fact:
\be
B^{\xi} \triangleright p \neq (B^{\xi}  \triangleright q) \oplus (B^{\xi}  \triangleright k)\,,
\ee
since in particular
\be
B^{\xi} \triangleright p_{1} = p_1 + \xi \left[\frac{\kappa}{2}\left(1-e^{-2p_0/\kappa}\right) - \frac{p_1^2}{2\kappa}\right] =   q_1+e^{-q_{0}/\kappa} k_{1} + \xi \left[\frac{\kappa}{2}\left(1-e^{-2(q_{0}+k_{0})/\kappa}\right) - \frac{(q_1+e^{-q_{0}/\kappa} k_{1})^2}{2\kappa}\right]
\ee

and
\begin{align}
 \left( (B^{\xi}  \triangleright q) \oplus (B^{\xi}  \triangleright k)\right)_{1} &= q_1 + \xi \left[\frac{\kappa}{2}\left(1-e^{-2q_0/\kappa}\right) - \frac{1}{2\kappa}q_1^2\right]\nonumber\\
&+e^{-q_{0}/\kappa}(1-\xi \frac{p_{1}}{\kappa}) k_1 +e^{-q_{0}/\kappa}\xi \left[\frac{\kappa}{2}\left(1-e^{-2k_0/\kappa}\right) - \frac{1}{2\kappa}k_1^2\right]  
\end{align}
are clearly different.

What does work to achieve covariance of the conservation equation is to account for a ``backreaction'' of the interacting momenta onto the boost rapidity $\xi$ \cite{Gubitosi:2013rna}. This is such that the rapidity with which the  momentum of the second outgoing particle of the vertex\footnote{The ordering refers to the sequence with which the momenta appear in the composition law.} transforms, is affected by the momentum of the first outgoing particle:
%\be
%\Lambda_\xi(q\oplus k) = \Lambda_\xi(q)\oplus\Lambda_{\xi\triangleleft\, q}(k)
%\ee
\be
B^{\xi} \triangleright p =  (B^{\xi}  \triangleright q) \oplus (B^{\xi\triangleleft q}  \triangleright k)\,,\label{eq:backreaction}
\ee
where $\xi\triangleleft q \equiv e^{-q_{0}/\kappa}\xi$ (a more general expression applies when considering finite transformations \cite{Gubitosi:2013rna}, however here we are only interested in the infinitesimal ones, that is the first order in $\xi$).
As discussed in detail in \cite{Amelino-Camelia:2013sba} such backreaction does not identify a preferred frame of reference and is fully compatible with relativistic invariance. 

We want to point out here a feature that will turn out to be very relevant for the results exposed in Section \ref{sec:RLrelativistic}. One can interpret the backreaction \eqref{eq:backreaction} in terms of a law of ``addition'' of boost generators that dictates how composed momenta transform\footnote{The fact that one might have to use such ``total boost" generator when transforming composed momenta in theories with nontrivial composition rules was suggested in \cite{AmelinoCamelia:2011yi, AmelinoCamelia:2011er, Kowalski-Glikman:2014wba}, but the link to the backreaction was not understood.}.  Namely, we may define the so-called ``total boost'' generator,
\begin{align}
N_{[q\oplus k]} = N_{[q]} + e^{-q_0/\kappa}N_{[k]},\label{eq:total_boost}
\end{align}
which is induced by the coproduct of the boost generator in the underlying Hopf algebra, Eq. \eqref{eq:kp_coalgebra}, in analogy with Eq. \eqref{eq:kP_composition_law} which defines the total momentum. Here the notation $N_{[q]}$ indicates that the relevant generator has nonzero brackets only with (i.e. acts on) $q$ and not with $k$:
\be
\begin{array}{lclclcl}
\left\{N_{[q]}, q_{0}\right\} &=&  q_{1}\,,&\qquad& \left\{N_{[q]}, q_{1}\right\} &=&  \frac{\kappa}{2}\left(1-e^{-2q_0/\kappa}\right) - \frac{1}{2\kappa}q_1^2\,, \\
\left\{N_{[q]}, k_{0} \right\}&=&  0\,,&\qquad& \left\{N_{[q]}, k_{1}\right\} &=&  0 \,,
\end{array}
\ee
and similarly for $N_{[k]}$. The ``total boost'' of rapidity $\xi$ the has the following action on the momenta of each of the two interacting particles:
\bea
q&\rightarrow& q+\xi \{N_{[q\oplus k]},q\}= q+\xi \{N_{[q]},q\}\,,\nonumber \\
k&\rightarrow& k+\xi \{N_{[q\oplus k]},k\}= k+\xi e^{-q_{0}/\kappa} \{N_{[k]},k\}= k+(\xi\triangleleft q) \{N_{[k]},k\}\,.
\eea
In other words: transforming each momentum with its own boost generator and incorporating the backreaction of the other momenta on the rapidity is completely equivalent to transforming each momentum instead with the total boost generator, without making reference to any backreaction.

One can explicitly check that the total boost action is also equivalent to the backreaction at the level of the sum $q\oplus k$ of the two momenta (remember that we work at first order in $\xi$):
\bea
( q\oplus k)_{0}+\xi \{N_{[q\oplus k]}, (q\oplus k)_{0}\} &=& q_{0}+  k_{0}+\xi \{N_{[q]}+e^{-q_{0}/\kappa} N_{[k]}, q_0+ k_{0}\}=q_{0}+  k_{0}+\xi \{N_{[q]}, q_0\}+\xi \{e^{-q_{0}/\kappa} N_{[k]},  k_{0}\} \nonumber\\
&=& B^{\xi}\triangleright q_{0}+  B^{\xi\triangleleft q}\triangleright k_{0} =\left(\left(B^{\xi}\triangleright q \right)\oplus \left( B^{\xi\triangleleft q}\triangleright k\right)\right)_{0}    \,,\nonumber\\ 
&&\nonumber\\
( q\oplus k)_{1}+\xi \{N_{[q\oplus k]}, (q\oplus k)_{1}\} &=& q_{1}+e^{-q_{0}/\kappa}  k_{1}+\xi \{N_{[q]}+e^{-q_{0}/\kappa} N_{[k]}, q_1+ e^{-q_{0}/\kappa} k_{1}\} \nonumber\\
%&=& q_{1}+e^{-q_{0}/\kappa}  k_{1}+\xi \{N_{[q]}, q_1+ e^{-q_{0}/\kappa} k_{1}\}+\xi \{e^{-q_{0}/\kappa} N_{[k]},  e^{-q_{0}/\kappa} k_{1}\}  \,,\nonumber\\
&=& q_{1}+e^{-q_{0}/\kappa}  k_{1}+\xi \{N_{[q]}, q_1\}-\frac{\xi}{\kappa}e^{-q_{0}/\kappa} \{N_{[q]},  q_{0}\}  k_{1}+\xi e^{-2q_{0}/\kappa}\{ N_{[k]},   k_{1}\}  \nonumber\\
&=&   (q_{1}+\xi \{N_{[q]} , q_1\}) + e^{-q_{0}/\kappa}(1-\frac{\xi}{\kappa} \{N_{[q]} , q_0\}) (k_{1}+\xi e^{-q_{0}/\kappa}\{N_{[k]} , k_1\})\nonumber\\
&=&(B^{\xi}  \triangleright q_{1}) + e^{-(B^{\xi}  \triangleright q_{0})/\kappa} (B^{\xi\triangleleft q}  \triangleright k_{1}) =  \left(\left(B^{\xi}  \triangleright q\right) \oplus \left(B^{\xi\triangleleft q}  \triangleright k\right)\right)_{1}\,.
\eea

To conclude this Subsection let us address the transformation rule of general interaction vertices, beyond the specific example  we used. The generalisation to a different number of incoming and outgoing particles is straightforward \cite{Gubitosi:2013rna}. The rapidity of each incoming (outgoing) particle receives a backreaction from the momenta of all the other incoming (outgoing) particles that come before it in the composition law (i.e. whose worldlines would be depicted above its own in a diagram of the kind of Figure \ref{fig:p_to_q_k}). Specifically, in the process
\be
p^{(1)}+...+p^{(n)}\to q^{(1)}+...+q^{(m)}\,,
\ee
the rapidity with which the particle with momentum $q^{j}$ is boosted is:
\be
\xi^{[q^{(j)}]}=\xi\triangleleft q^{(1)}\triangleleft ... \triangleleft q^{(j-1)}\equiv \xi\triangleleft (q^{(1)}\oplus ... \oplus q^{(j-1)})\,,
\ee
and similarly the rapidity with which the particle with momentum $p^{k}$ is boosted is:
\be
\xi^{[p^{(k)}]}=\xi\triangleleft p^{(k)}\triangleleft ... \triangleleft p^{(k-1)}\equiv \xi\triangleleft (p^{(1)}\oplus ... \oplus p^{(k-1)})\,.
\ee
Again, we can interpret the backreaction in terms of the action of a total boost. Boosting each momentum with its own boost generator, $N_{[p^{(i)}]}$, and incorporating the backreaction on the rapidity, as explained above, is completely equivalent to boosting each momentum of the incoming particles with the total boost generator 
\begin{align*}
N_{[\bigoplus_{i=1}^n p^{(i)}]} = N_{[p^{(1)}]} + e^{-p^{(1)}_0/\kappa}N_{[p^{(2)}]} + e^{-\left(p^{(1)}_0+ p^{(2)} _0\right)/\kappa}N_{[p^{(3)}]} + \dots + e^{-\left(\sum_{i=1}^{n-1} p^{(i)}_0\right)/\kappa}N_{[p^{(n)}]} \,,
\end{align*}
and similarly for the outgoing particles.
\subsection{Multiple interaction vertices}
\label{sub:kPmultivertex}

In systems of multiple vertices that are causally connected (i.e. that share  the worldline of at least one particle) translations affect not just the composition of momenta, but  they also link observers located at the different interaction events. One  then expects to see an interplay between the relative locality effects (affecting how distant observers see interactions) and deformed Lorentz transformations \cite{AmelinoCamelia:2011yi}. 

\begin{figure}[h]
\begin{center}
\includegraphics[width=0.6\textwidth]{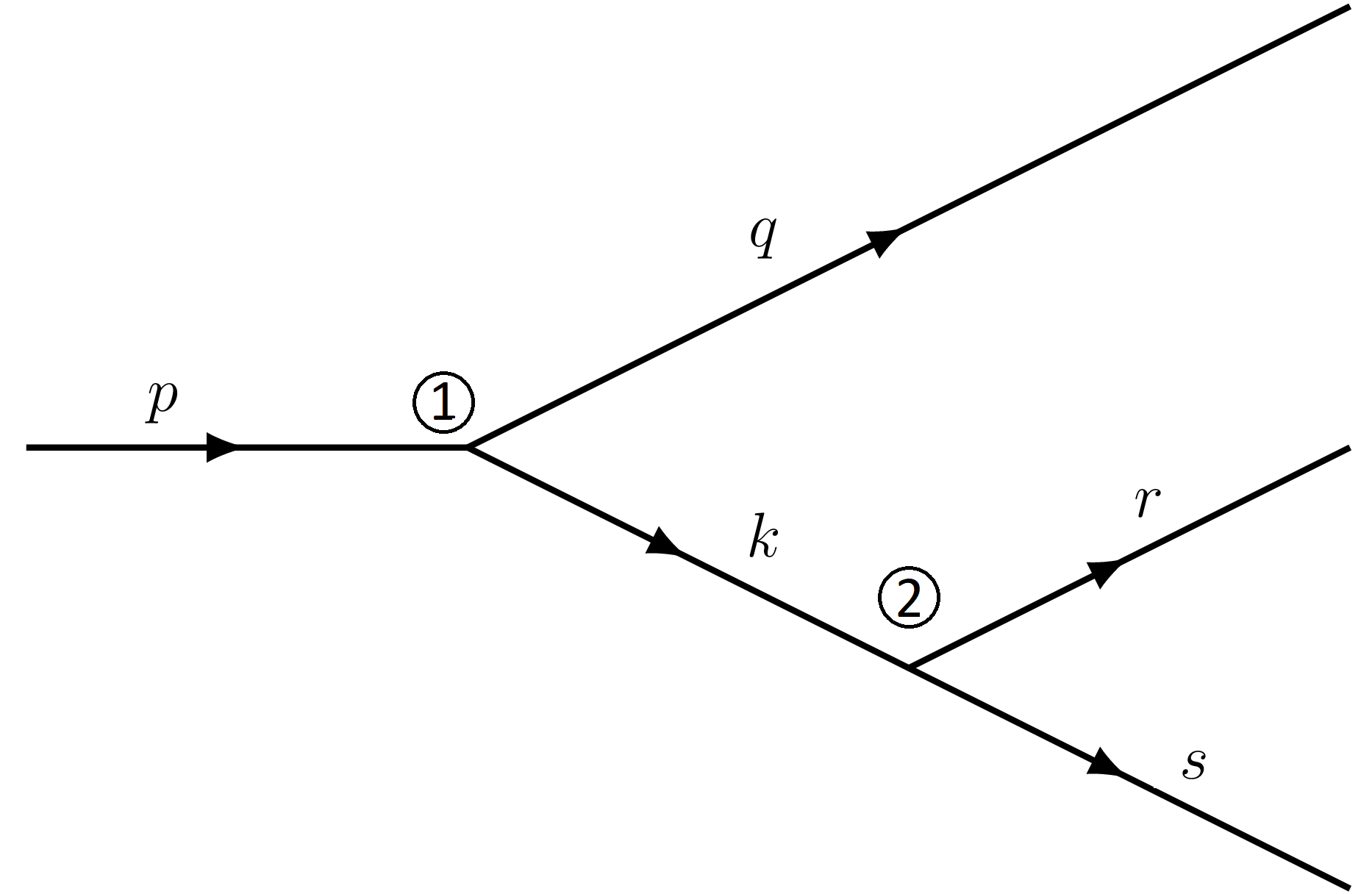}
\caption{Interaction diagram with multiple vertices. Vertex number $1$ has one incoming particle with momentum $p$ and two outgoing particles, with momenta $q$ and $k$. The particle with momentum $k$ then decays at vertex number $2$ into two particles, with momentum $r$ and $s$ respectively.}
\label{fig:problem.diagram}
\end{center}
\end{figure}

As for the single-vertex case, we discuss this using a specific physical example, represented in Figure \ref{fig:problem.diagram}: in the event $1$ a particle with momentum $p$ decays into two particles, with momentum $q$ and $k$ respectively (again note that the ordering is important). Then the particle with momentum $k$ undergoes a decay  (event number $2$) into two particles with momentum $r$ and $s$ respectively. Basically, we are gluing a second interaction event to the process discussed in the previous Subsection.
The conservation rules of momenta at the two vertices read: 
\bea
p_0 &=& q_0 + k_0\,,\qquad p_1 = q_1 + e^{-q_0/\kappa}k_1\,, \\ 
k_0 &=& r_0 + s_0\,,\qquad k_1 = r_1 + e^{-r_0/\kappa}s_1\,.
\eea

It is then immediate to see how the interplay between Lorentz transformations and translations can produce apparently paradoxical results.
In fact, if we  perform a boost with rapidity $\xi$  by naively applying  the procedure discussed in the previous Subsection to the two vertices separately  we  get  that vertex $1$ transforms as described in the  one-vertex example:
\be
B^{\xi} \triangleright p =  (B^{\xi}  \triangleright q) \oplus (B^{\xi\triangleleft q}  \triangleright k)\,,
\ee
and similarly   vertex $2$  transforms as:
\be
B^{\xi} \triangleright k =  (B^{\xi}  \triangleright r) \oplus (B^{\xi\triangleleft r}  \triangleright s)\,.
\ee
This  is clearly inconsistent, since the momentum  $k$ is boosted with different rapidities at the two interaction events ($\xi\triangleleft q$ in the first case and simply $\xi$ in the second case), while the translation that connects the two endpoints of the worldline of the particle with momentum $k$ does not change the value of the momentum, $k\rightarrow T_{a}\triangleright k = k$, which should then be seen as constant along the worldline also by a boosted observer.  So this paradox can be traced back to the fact that in performing the boost transformation in this way one is neglecting the properties of the translation transformation that links the two endpoints of the worldline of the particle that connects the two interaction events. 
\subsubsection{Interlude: on the composition of deformed symmetry transformations}
In order to better understand   the interplay between boosts and translations let us characterize it by inspecting how these transformations are composed.
As done before, let us denote a translation with parameter $a=(a^{0},a^{1})$ as $T_a$ and a boost with  rapidity $\xi$ as $B^{\xi}$. One can show that the composition of two translations gives
\be
T_a\circ T_b =T_b\circ T_a= T_{a+b}\,,
\ee
while the composition of two boosts gives
\be
B^{\xi}\circ B^{\zeta}=B^{\zeta}\circ B^{\xi} =B^{\xi+\zeta}\,,
\ee
in both cases just as in special relativity. The commutativity of the translation generators $P_{\mu}$ and that of the boost generators $N$ implies the commutativity of the composition of two translations or two boosts\footnote{Remember that we are working in $1+1$ dimension, so that there is just one boost generator. Of course in higher dimensions boosts in different directions do not commute and generate a Thomas-Wigner rotation.}.
Departures from special relativity are seen when studying distant relatively boosted observers, related by the composition of a  boost and a translation. In special relativity one finds that 
\be
T_a\circ B^{\xi} = B^{\xi}\circ T_{a'}\,, \qquad a' = B^{\xi}\triangleright a\,.\label{eq:TBcommutation}
\ee 
That is, if in a given inertial frame the two extremes of a worldline are related by a translation with parameter $a$, in another frame, relatively boosted with respect to the first one with rapidity $\xi$, the two extremes of the worldline are related by a translation with boosted parameter $a'$, which is simply a function of the transformation parameters $a$ and $\xi$:
\begin{align}
{a}'(p)^0&= a^0 + \xi a^1\,, \\
{a}'(p)^1&= a^1 + \xi a^0\,. \label{eq:momentumdependenttranslation}
\end{align}
In other words, the translation parameter linking two distant relatively boosted observers depends on whether the boost or the translation is performed first.
This can be ascribed to the fact that in special relativity the commutator between a boost and a translation gives a translation \cite{AmelinoCamelia:2011cv}. 

In the $\kappa$-Poincar\'e model however one finds
\be
T_a \circ B^{\xi} = B^{\xi} \circ T_{a'(p)} \,,
\ee
where the new ``translation parameter'' $a'(p)$ depends on the momentum of the worldline whose ends were related by the translation $T_{a}$ in the non-boosted frame: 
\begin{align}
{a}'(p)^0&= a^0 + \xi e^{-2p_0/\kappa}a^1\,, \\
{a}'(p)^1&= a^1 - \xi\left(\frac{p_1 a^1}{\kappa} - a^0\right)\,.
\end{align}
This is due to the fact that the commutator between the \kp boost and  translation generators gives a nonlinear function of the translation generators,  Eq. \eqref{eq:kp_algebra}. So  if in a given inertial frame the two extremes of a worldline\footnote{In principle at this point of our analysis we have not yet defined worldlines, since we only have a momentum space construction. Spacetime (and worldlines) will be defined in Section \ref{sec:RL}, after which we will come back to this discussion to make it more precise.} are related by a translation with parameter $a$ then in another frame, relatively boosted with respect to the first one with rapidity $\xi$, the two extremes of the worldline are no longer related by a pure translation.  This statement will be given a more precise physical characterization at the end of Section \ref{sec:RLrelativistic}, once a fully covariant definition of spacetime in the presence of interactions will have been provided, so as to be able to describe finite worldlines. 
For the moment let us conclude this interlude by observing that what we have just exposed is a mixing between momenta and translation parameters generated by the composition of translations and boosts. This can be ascribed to the fact that actually  \kp symmetry transformations mix the phase space coordinates (as opposed to special relativistic transformations that do not mix spacetime coordinates and momenta).  To show this explicitly we first need to construct the spacetime associated to the \kp momentum space, which we do in Section \ref{sec:RL}.

\subsubsection{Consistent transformation rules of multi-interactions systems}
Going back to the analysis of  the multi-interactions system of Figure \ref{fig:problem.diagram}, we find that the consistent way of acting with a boost transformation, which  preserves relativistic invariance, is to boost each particle with a rapidity that receives  a backreaction from the total momentum of all the causally connected particles that come before that one in the momenta composition rule (i.e. whose lines are drawn above the one of the given particle in the convention adopted for our figures). In particular, one has to account also for the particles that do not interact directly with the  particle whose momentum is being boosted. In the specific example of the interactions depicted in Figure \ref{fig:problem.diagram}, this amounts to the following:
\bea
p&\to& B^{\xi} \triangleright p\,, \nonumber\\
q&\to& B^{\xi} \triangleright q\,, \nonumber\\
k&\to& B^{\xi\triangleleft q} \triangleright k\,, \\
r&\to& B^{\xi\triangleleft q} \triangleright r  \,, \nonumber\\
s&\to& B^{\xi\triangleleft q\triangleleft r} \triangleright s = B^{\xi\triangleleft (q\oplus r)} \triangleright s\,. \nonumber
\eea

One can easily verify that this law of transformation preserves the conservation of momenta at each interaction vertex:
\bea
B^{\xi} \triangleright p &=&  (B^{\xi}  \triangleright q) \oplus (B^{\xi\triangleleft q}  \triangleright k)\,,\nonumber\\
B^{\xi\triangleleft q} \triangleright k &=&  (B^{\xi\triangleleft q}  \triangleright r) \oplus (B^{\xi\triangleleft (q\oplus r)}  \triangleright s)\,,
\eea
and that no inconsistencies of the sort described at the beginning of this Subsection arise. One can also show that these transformation rules can  be written in terms of the ``total boost'' defined in the previous Subsection, where the total boost of course accounts for the total momentum of the system, $p=q\oplus k=q\oplus r\oplus s$, as follows:
\bea
 p&\rightarrow&  p+\xi \{N_{[p]}, p\}\,,\nonumber\\
q&\rightarrow&q+\xi \{N_{[q\oplus k]}, q\}=q+\xi \{N_{[q]}+e^{-q_{0}/\kappa} N_{[k]}, q\}=q+\xi \{N_{[q]}, q\}\,,\nonumber\\
  k&\rightarrow&  k+\xi \{N_{[q\oplus k]}, k\}=k+\xi \{N_{[q]}+e^{-q_{0}/\kappa} N_{[k]},  k\}=k+\xi e^{-q_{0}/\kappa} \{ N_{[k]},  k\}\,,\nonumber\\
  r&\rightarrow& r+\xi \{N_{[q\oplus r\oplus s]}, r\}=r+\xi \{N_{[q]}+e^{-q_{0}/\kappa} N_{[r]}+e^{-(q_{0}+r_{0})/\kappa} N_{[s]}, r\}=r+\xi e^{-q_{0}/\kappa} \{ N_{[r]}, r\}\,,\nonumber\\
   s&\rightarrow& s+\xi \{N_{[q\oplus r\oplus s]}, s\}=s+\xi \{N_{[q]}+e^{-q_{0}/\kappa} N_{[r]}+e^{-(q_{0}+r_{0})/\kappa} N_{[s]},  s\}=s+\xi e^{-(q_{0}+r_{0})/\kappa}  \{N_{[s]},  s\}\,.
\eea

The way boost transformations act on systems of interactions  shows what can be thought of as an effect of the nonlocal features of the model, since  the action of a boost on one of the vertices belonging to an interaction chain depends on the particle content and interaction structure  of the whole system. However, as long as an observer has  access to one vertex only, she can just use the one-vertex boost rule without finding any inconsistency. In fact, for the example of Figure \ref{fig:problem.diagram}, an observer that has only access to vertex $2$ would simply see a rescaled rapidity  $\tilde\xi=\xi\triangleleft q $.  It is only when the observer has access to information about both of the vertices (e.g. because she exchanges information with an observer local to the other vertex), that the ``backreaction from the total momentum'' rule becomes relevant.

\subsubsection{A selection rule for physically allowed interactions}

\begin{figure}[h]
\begin{center}
\includegraphics[width=0.8\textwidth]{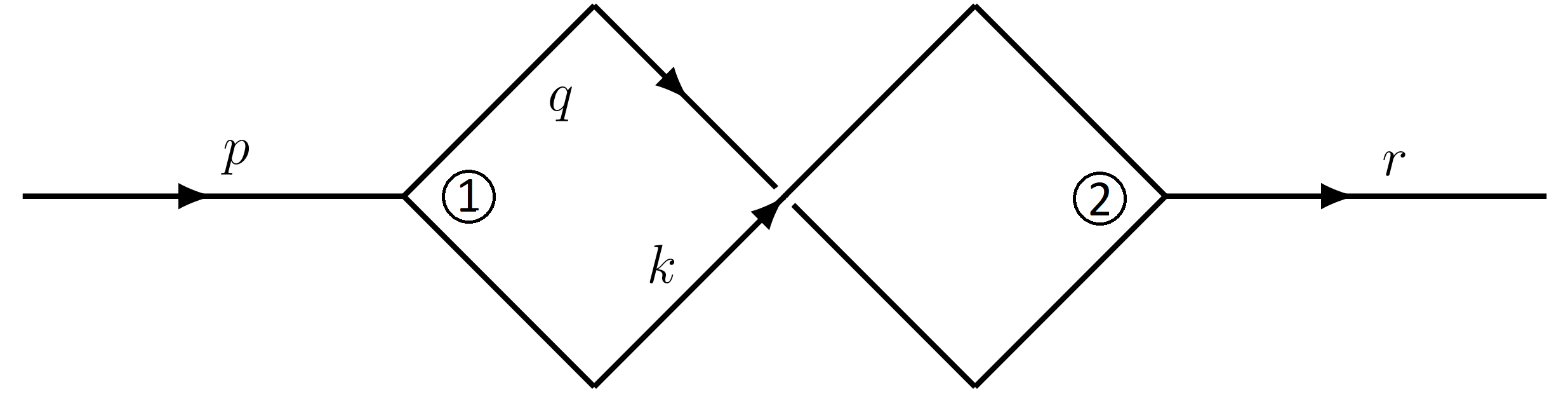}
\caption{Interaction diagram with two interaction vertices and crossing lines. At vertex number $1$ the particle with momentum $p$ decays into two particles with momentum $q$ and $k$ respectively. These same particles interact at vertex $2$, but in inverted order, to create the particle with momentum $r$. We remind the reader that the order of (noncommutative) summation of momenta is encoded in the diagram in such a way that particles enter into the sum following the ordering of their lines in the diagram from top to bottom. In this specific diagram, at vertex $1$ the momenta of the two outgoing particles are summed as $q\oplus k$, while at vertex $2$ the momenta of the incoming particles are summed as $k\oplus q$.}
\label{fig:transl.crossing.lines}
\end{center}
\end{figure}

Clearly the boost transformation defined above is only consistent for interaction diagrams that do not have crossing lines, so that  the ordering of the momenta is always well defined (remember that the crossing of lines implies that the ordering in which the momenta are added is inverted).  For example, in the process depicted in Figure \ref{fig:transl.crossing.lines}  the conservation law of momenta at the two vertices read 
\bea
p&=&q\oplus k\,,\nonumber\\
r&=&k\oplus q \,. \label{eq:crossing}
\eea 
According to the procedure defined above, a boost with rapidity $\xi$ would act on the first vertex as:
\bea
p&\to& B^{\xi} \triangleright p\,, \nonumber\\
q&\to& B^{\xi} \triangleright q\,, \nonumber\\
k&\to& B^{\xi\triangleleft q} \triangleright k\,, 
\eea
while the second vertex would transform as
\bea
r&\to& B^{\xi} \triangleright r\,, \nonumber\\
k&\to& B^{\xi} \triangleright k \,, \nonumber\\
q&\to& B^{\xi\triangleleft k} \triangleright q\,, 
\eea
so that an inconsistency appears in the transformation rules of the momenta $q$ and $k$. The fact that it is not possible to define a boost symmetry transformation provides motivation to exclude this kind of diagrams from those that are physically allowed. And indeed it  turns out that  diagrams with crossing lines also violate conservation of total momentum, as can be seen by comparing the first and second lines of Eq. \eqref{eq:crossing}, from which it follows that $p\neq r$. This is  a sign that also translation invariance is broken by these diagrams (see also \cite{Amelino-Camelia:2014qaa}).

 \section{From momentum space to phase space - the relative locality framework}\label{sec:RL}
 
 In the previous Section we have analyzed the relativistic properties of the \kp model, which is in principle a model for momentum space only. However some of its phenomenological applications actually require that the associated spacetime is constructed. This is for example the case with   studies of the time of flight of particles from astrophysical sources, where one looks for  a difference in the arrival time of particles with different energies \cite{AmelinoCamelia:2008qg}. In this Section we review a framework that was proposed for this purpose, known as relative locality \cite{AmelinoCamelia:2011bm,AmelinoCamelia:2011pe}. The framework is in principle suited to any model with nontrivial momentum space properties. Here we  focus on the specific case where the momentum space is that of the \kp model and  review how the full phase space of the \kp relative locality model is constructed for a single free particle and in the more complex  case of interacting particles. The following Section \ref{sec:RLrelativistic} is then devoted to study whether the relative locality phase space construction is compatible with the relativistic symmetries of the \kp model.

\subsection{Free particles}
The definition of spacetime in the case of one  free particle is conceptually relatively simple, because the particle lives on one point of the curved momentum manifold, say $p$.\footnote{Of course relatively boosted observers associate a different point on momentum space to the same particle. In this sense the spacetime constructed by the relative locality framework is observer-dependent \cite{AmelinoCamelia:2011bm}.} Then one can take the momentum space as the base manifold and construct spacetime as the cotangent space of the momentum manifold at the point $p$. This is a completely analogous construction to the one of general relativity where  momentum space is the cotangent space of the spacetime manifold at a point in spacetime. In this way one can define the free particle dynamics in a canonical way, with the role of spacetime and momentum space exchanged with respect to the usual construction.  

One starts by defining the spacetime coordinates $x^\mu$ as canonically conjugated to momenta via Poisson brackets,\footnote{One can in principle make a different choice for the symplectic structure, since the physical content of the model is not affected by this choice \cite{AmelinoCamelia:2011nt}.}
\be
\{x^u,p_\nu\} = \delta^\mu_\nu\,.
\ee

Then the dynamics  of a single free particle is  described  by the following action:
\be
\label{eq:free_action}
S^{\text{free}} = \int \D \lambda \left( -x^\mu \dot p_\mu + \mathcal N \left( D(p)^2 - m^2\right) \right)\,.
\ee
The over-dot indicates a derivation with respect to the affine parameter $\lambda$. The parameter $\mathcal N$ is a Lagrange multiplier enforcing the on-shell relation, $D(p)^2-m^2=0$, where $D(p)$ is the geodesic distance in momentum space of $p$ from the origin of the manifold\footnote{Some care is required when defining the point $0$ in momentum space, since in a general curved manifold there is no such preferred point. Defining $0$ by requiring it to have coordinate expression $0$ is highly ambiguous because of the many possible coordinate charts. A better way of defining $0$ is possible in case the coordinate functions on the momentum manifold are elements in a Hopf algebra. The co-unit $\epsilon$ of this Hopf algebra then defines the coordinates of the origin via $\epsilon(P_\mu) = P_\mu (0)$. This is coordinate independent, since a change of coordinates can be related to a basis change in the Hopf algebra. For the \kp model, the comoving coordinates of the origin all vanish.}, and $m$ is the mass of the particle. 
Varying the action \eqref{eq:free_action} with respect to $x^\mu$ yields conservation of momentum along the worldline,
\begin{align}
\dot p_\mu = 0\,,
\end{align}
while variation with respect to the momentum $p_\mu$ yields the evolution equation for the spacetime coordinates:\footnote{Since $\mathcal N$ behaves identically as a multiplicative factor in each component of $\dot x^\mu$, it does not affect the worldlines of particles and its value constitutes nothing more than a convention for the normalization of the momenta.}
\begin{align}
\label{eqn:xdot}
\dot x^\mu = - \mathcal N \frac{\partial \mathcal C}{\partial p_\mu}\,,
\end{align}
where $ \mathcal C(p) \equiv D(p)^2 - m^2$.
When the momentum geometry is that of  the \kp model (i.e. that of a de Sitter manifold), the equations of motion and the constraint equations take the explicit form \cite{Gubitosi:2013rna}:
\bea
m &=& \kappa \,\arccosh \left( \cosh\frac{p_0}{\kappa}-e^{\frac{p_0}{\kappa}}\frac{(p_1)^2}{2\kappa^2}\right)\,, \label{eq:rel_loc_disp_rel}\\
\dot p_\mu &=& 0\,, \label{eq:rel_loc_const_p} \\
\frac{\partial x^1}{\partial x^0} &\equiv& \frac{\dot x^1}{\dot x^0} = \frac{2\kappa p_1}{\kappa^2\left(e^{-2 p_0/\kappa}-1\right) + (p_1)^2}\,. \label{eq:rel_loc_worldline}
\eea
Integrating the last equation and using the on-shell relation  \eqref{eq:rel_loc_disp_rel} one finds the worldline of a \kp particle:
\be
x^{1}(x^{0}) = x^{1}(0) + v(p)x^{0},\qquad v(p) =\frac{e^{p_0/\kappa}\sqrt{e^{2 p_0/\kappa}+1-2\,  e^{p_0/\kappa}\cosh(m/\kappa)}}{1-e^{p_0/\kappa}\cosh(m/\kappa)}\,. \label{eq:rel_loc_worldline1}
\ee
Note that \eqref{eq:rel_loc_disp_rel} is different from the dispersion relation derived directly from the \kp Casimir,  Eq. \eqref{eq:dispersion_relation}. This is because in the context of the \kp model $D(p)$ turns out to be a function of the representation in momentum space of the  \kp Casimir.  Asking that the two on-shell relations are equivalent simply amounts  to a redefinition of the physical mass $m$.  
Finally, because the momentum $p_\mu$ is a constant of motion,  the velocity $v(p)$ is constant as well. The linearity of the worldlines with respect to spacetime coordinates may be interpreted by saying that the spacetime we have constructed is flat. The deformed expression for $v(p)$ can be attributed to the fact that momentum space, on the other hand, has a nontrivial geometry.

\subsection{Interacting particles}

 When more than one particle is at play, the construction outlined above requires us to build a different set of spacetime coordinates $x_I^\mu$, each living on the cotangent space of the momentum manifold at a different point $p^I$, corresponding to the momentum of particle $I$. If the particles are non interacting, then one can simply write down the total action as the sum of the free actions of each particle, labelled with index $I$:
 \bea
S^{tot}&=&\sum_I S_I^{\text{free}}\,, \nonumber \\
S_I^{\text{free}} &=& \int_{-\infty}^{\infty} \D \lambda \left( -x_I^\mu \dot p^I_\mu + \mathcal N_I \left( D(p^I)^2 - m_I^2\right) \right)\,, \label{eq:free_action_multiparticle}
\eea
so that for each $I$ spacetime coordinates $x_{I}^{\mu}$ are canonically conjugate to the momenta $p_{\mu}^{I}$.
 If the particles are interacting the issue  of how to define spacetime at the interaction point arises. In particular, how to indicate that the particles are actually interacting? Since the spacetime coordinates we have defined for each particle in the noninteracting case live in different cotangent spaces, it does not make sense to ask that the coordinates $x_I^\mu$ take the same value for all $I$'s at the interaction event. The solution provided within the relative locality framework  is to introduce a boundary interaction term in the action, with a constraint that enforces momentum conservation at the interaction \cite{AmelinoCamelia:2011bm}. Thus the total action for $n$ incoming and $m$ outgoing particles reads:
  \bea
S^{tot}&=&\sum_{I=1}^{n+m} S_I^{\text{free}}+S^{int} \,, \nonumber \\
S_I^{\text{free}} &=& \pm\int_{\lambda_{0}^{I}}^{\pm \infty} \D \lambda \left( -x_I^\mu \dot p^I_\mu + \mathcal N_I \left( D(p^I)^2 - m_I^2\right) \right)\,, \nonumber\\
S^\text{int} &=& z^\mu \mathcal K_\mu(p_1(\lambda_0^{1}),\dots,p_n(\lambda_0^{n}), p_{n+1}(\lambda_0^{n+1}),\dots,p_m(\lambda_0^{m}))\,, \label{eq:rel_loc_action_total}
\eea
where the upper (lower) signs are for  outgoing (incoming) particles, $\lambda_0^{I}$ is the value of the affine parameter at the endpoint of the worldline of each particle where the interaction occurs, and $z^\mu$ is a Lagrange multiplier enforcing the conservation law  $\mathcal K_\mu(p_1(\lambda_0^{1}),\dots,p_n(\lambda_0^{n}), p_{n+1}(\lambda_0^{n+1}),\dots,p_m(\lambda_0^{m}))=0$. \\

Within the \kp model, the conservation law accounts  for the deformed composition of momenta:
\be
p_1\oplus \dots \oplus p_n= p_{n+1}\oplus \dots \oplus p_m\,. \label{eq:conservation_law}
\ee 
In principle there are several possibilities for the actual form of $\mathcal K_\mu$, all compatible with \eqref{eq:conservation_law}. We choose to  write the boundary term as the difference between the total momentum before and after the interaction:
\be
\mathcal K_\mu = p_1\oplus \dots \oplus p_n - (p_{n+1}\oplus \dots \oplus p_m)\,.\label{eq:rel_loc_conservation}
\ee
In fact, this choice is  consistent with translational invariance in multi-interactions systems \cite{AmelinoCamelia:2011nt}, as  is reviewed in the following Section. Note that with this choice $\mathcal K_\mu$ does not transform as a vector. For this reason the way Lorentz invariance of interactions is realized  turns out to be nontrivial, as we show in Section \ref{sec:RLrelativistic}.  Had we used the deformed antipode operator $\ominus$ instead of the simple minus sign then $\mathcal K_\mu$ would have been covariant. However that choice would have spoiled translational invariance of multi-interaction systems.

 From varying the action one gets  similar constraints for each interacting particle as those found for the free particle, Eqs. \eqref{eq:rel_loc_disp_rel}-\eqref{eq:rel_loc_worldline}. Additionally, the interaction term yields an additional constraint on the endpoints of the worldlines at the interaction,
\be 
x^\mu_I(\lambda_0^{I}) = \mp z^\nu \frac{\partial \mathcal K_\nu }{\partial p^I_\mu }\bigg|_{\lambda=\lambda_0}\,,\label{eq:rel_loc_constraint_interaction}
\ee
where the upper (lower) sign is for outgoing (incoming) particles. In special relativity one has $\mathcal K_\mu = p_1+ \dots + p_n - (p_{n+1}+ \dots + p_m)$ and the worldlines of all particles  end at $x^\mu_I=z^\mu$, so that the interaction is local. In this case however one  easily sees that only if the interaction happens at $z^\mu=0$ then all worldlines end at $x^\mu_I=z^\mu=0$. If instead $z^\mu\neq 0$ then each worldline  ends at a different value of $x^\mu_I$, because of the nonlinearity of the composition law of momenta, such that in general $ \frac{\partial \mathcal K_\nu }{\partial p^I_\mu }\neq  \frac{\partial \mathcal K_\nu }{\partial p^J_\mu }$ . This is a manifestation of relative locality: only a local observer, $z^\mu=0$, sees the interaction as local, while other observers, $z^\mu\neq 0$, see each worldline ending at a slightly different point. To make this statement more precise one needs to specify the rules of transformation between observers and how spacetime symmetries act on worldlines and interaction events. This is the focus of the following Section.

\section{Relativistic analysis of the $\kappa$-Poincar\'e relative locality spacetime}\label{sec:RLrelativistic}

Having defined a full phase space for the \kp model in the previous Section, here we set to the task of demonstrating that it is indeed a relativistically viable, i.e. compatible with the symmetries of the \kp algebra, in the same way as the momentum space \kp model is. 

We mentioned in the previous Section that translational invariance of the \kp relative locality model has already been established \cite{AmelinoCamelia:2011nt}. We review the analysis in the following and complement it with the demonstration of boost invariance, which was until now a missing ingredient except that in the simple free particle case. The presence of interactions raises several difficulties, related to the interplay between deformed translation and boost transformations. For example, as we discussed briefly at the end of the previous Section, the constraint enforcing momentum conservation and compatible with translational invariance, Eq. \eqref{eq:rel_loc_conservation}, does not transform covariantly under boosts. This raises well grounded doubts on the relativistic invariance of the  interaction constraints in Eq. \eqref{eq:rel_loc_constraint_interaction}. 

To demonstrate the invariance of the relative locality \kp model we follow  a similar line of reasoning  as that used in the analysis of the relativistic properties of the \kp momentum space in Section \ref{sec:symmetries_kp_momentum_space}. We first discuss the free particle model, then proceed to one interaction vertex and we finally look at multiple interactions.

\subsection{Free particle}

To demonstrate the relativistic invariance of the \kp relative locality model with one  free particle, we have to inspect the equations of motion and constraints encoded in Eqs. \eqref{eq:rel_loc_disp_rel}-\eqref{eq:rel_loc_worldline}. The first equation is nothing else than the dispersion relation, whose  relativistic invariance was already discussed in Section \ref{sec:symmetries_kp_momentum_space}. The second equation is trivially invariant under translations and boosts, since these transformations are closed with respect to momenta, see Eqs. \eqref{eq:boosted_momenta} and \eqref{eq:translated_momenta}. So we are only left with showing the invariance of the worldline.\footnote{For a complete discussion of the relative locality effects that emerge already at this level of complexity we refer the reader to \cite{Amelino-Camelia:2013uya,AmelinoCamelia:2011nt,Barcaroli:2015eqe}.}

For a free particle the identification of the symmetry generators and their action on spacetime coordinates is straightforward. The translation generators $P_{\mu}$ are identified with the particle's momentum charge $p_{\mu}$  and act on coordinates as:
\be
x^{\mu}\to T_{a}\triangleright x^{\mu}\equiv x^{\mu}+a^{\nu}\{p_{\nu},x^{\mu}\}=x^{\mu}-a^{\mu}\,. \label{eq:coordinate_translation}
\ee
To find find how coordinates transform under Lorentz  transformations one starts by observing that the  boost generator can be represented on the phase space of a particle as follows: 
\be
\label{eq:boost_generator_representation}
N = p_1 x^0 + x^1 \left(\frac{\kappa}{2}\left(1-e^{-2 p_0/\kappa}\right)-\frac{ (p_1)^2}{2\kappa}\right)\,,
\ee
so that it closes the algebra \eqref{eq:kp_algebra} with translation generators.
Then its action over coordinates reads:
\bea
x^0\to  B^{\xi}\triangleright x^{0} \equiv x^0 +  \xi\{N,x^0\} = x^0 - \xi e^{-2p_0/\kappa}x^1, \nonumber\\
x^1\to  B^{\xi}\triangleright x^{1} \equiv x^1 +  \xi\{N,x^1\} = x^1 + \xi\left(\frac{p_1x^1}{\kappa} - x^0\right).\label{eq:boosting_coordinates}
\eea

\subsubsection{Invariance under translations}

Under translations the worldline \eqref{eq:rel_loc_worldline1} transforms as:
\be
x^{1} = \bar x^{1} + v(p) x^{0}\to x^{1}-a^{1} =\bar x^{1} + v(p) (x^{0}-a^{0})\,.
\ee
This is a covariant transformation, since the translated worldline can be written in the same form as the original worldline,
\be
x^{1} =\bar{\bar{x}}^{1} + v(p) x^{0}\,,
\ee
upon a rescaling of the integration constant,  $\bar{ \bar{x}}^{1}= \bar x^{1} +a^{1} - v(p) a^{0}$.

\subsubsection{Invariance under boosts}
Under boosts the worldline  \eqref{eq:rel_loc_worldline1} transforms as: 
\be
x^{1} = \bar x^{1} + v(p) x^{0}\rightarrow  x^{1} + \xi\left(\frac{p_1x^1}{\kappa} - x^0\right) =\bar x^{1} + v(p') (x^{0}- \xi e^{-2p_0/\kappa}x^1)\,,\label{eq:free_worldline_boost}
\ee
with $v(p') = v(p)+\xi \{N,v(p)\}$. Then covariance under boosts is encoded by the  statement that the equality on the right hand side is verified if and only it the equality on the left hand side is. This can be verified by taking into account the dispersion relation \eqref{eq:rel_loc_disp_rel} and remembering that we focus on infinitesimal transformations, i.e. first order in $\xi$.

\subsection{Interacting model: one  vertex}

In Section \ref{sec:symmetries_kp_momentum_space} we demonstrated the invariance of the conservation law of momenta in interactions, which involved the introduction of a backreaction of momenta on the boost rapidity. Once the full phase space is constructed, one also needs to check the covariance of the particles' worldlines and of their endpoints, i.e. of the boundary condition \eqref{eq:rel_loc_constraint_interaction}. This entails establishing the transformation rules of the interaction parameter $z^\mu$ besides those of the coordinates of all particles intervening in the interaction. Because it will turn out that in order for the interaction to be covariant the transformation rules of coordinates are nontrivial, we  first focus on this aspect and then show that the symmetry transformations also leave the worldlines invariant.

We start by  considering as an example the same vertex as in Section \ref{sec:symmetries_kp_momentum_space}, depicted in Figure \ref{fig:p_to_q_k}. For this vertex the momentum conservation constraint function $\mathcal K_\mu$, Eq. \eqref{eq:rel_loc_conservation}, reads:
\bea
\mathcal K_0&=& p_0 - (q_0 + k_0)\,,\nonumber\\
\mathcal K_1&=&  p_1 - ( q_1 + e^{-q_0/\kappa}k_1)\,.
\eea
We call $x^\mu$, $y^\mu$, $w^\mu$ the spacetime coordinates dual to $p_\mu$, $q_\mu$, $k_\mu$ respectively (such that $\{x^\mu,p_\nu\}=\{y^\mu,q_\nu\}=\{w^\mu,k_\nu\}=\delta^\mu_\nu$). The boundary conditions \eqref{eq:rel_loc_constraint_interaction} then read (for notational simplicity we leave the dependence on $\lambda_{0}^{I}$ implicit):
\bea
x^0&=& z^0\,, \label{eq:x0_constraint}\\
x^1&=& z^1 \,,\\
y^0&=& z^0-z^1 e^{-q_0/\kappa}\frac{k_1}{\kappa} \,,\\
y^1&=& z^1\,,\\
w^0&=& z^0\,,\\
w^1&=&  z^1 e^{-q_0/\kappa} \label{eq:w1_constraint}\,.
\eea
In this explicit example one clearly sees that the interaction coordinates only coincide if $z^\mu=0$ (for which $x^{\mu}=y^{\mu}=w^{\mu}=0$), i.e. for local observers, while distant observers for which $z^{\mu}\neq 0$ (see Eq. \eqref{translationz}) see the interaction as nonlocal.

\subsubsection{Invariance under translations}

The issue of whether the relative locality framework is invariant under translations was studied in \cite{AmelinoCamelia:2011nt, Amelino-Camelia:2014qaa}. Here we briefly review the results in order to establish the formalism that will turn useful also in the analysis of Lorentz transformations and to adapt the computations to our convention for the Poisson brackets between coordinates and momenta.  

What was shown in \cite{AmelinoCamelia:2011nt} is that the boundary condition \eqref{eq:rel_loc_constraint_interaction} is invariant under the translations generated by the total momentum charge acting on the coordinates. In the example we are considering:
\bea
T_{a}\triangleright x^{\mu}&\equiv &x^{\mu}+a^{\nu}\{p_{\nu},x^{\mu}\}=x^{\mu}-a^{\mu}\,,\\
T_{a}\triangleright y^{\mu}&\equiv &y^{\mu}+a^{\nu}\{(q\oplus k)_{\nu},y^{\mu}\}=y^{\mu}-a^{\mu}+\delta^\mu_0a^1e^{-q_0/\kappa}\frac{k_1}{\kappa}\,,\\
T_{a}\triangleright w^{\mu}&\equiv &w^{\mu}+a^{\nu}\{(q\oplus k)_{\nu},w^{\mu}\}=w^{\mu}-a^{0}\delta^\mu_0-a^{1}e^{-q_0/\kappa}\delta^\mu_1\,.
\eea
Note that the spacetime coordinates associated to different particles transform differently under translations, and in particular coordinates $x^{\mu}$ transform classically, while coordinates $y^{\mu}$ and $w^{\mu}$ do not.
Applying these transformations to the constraint equations \eqref{eq:x0_constraint}-\eqref{eq:w1_constraint} one finds:
\bea
x^0-a^0&=& T_{a}\triangleright z^0\,, \\
x^1-a^1&=& T_{a}\triangleright z^1 \,,\\
y^0-a^0+a^1e^{-q_0/\kappa}\frac{k_1}{\kappa}&=& T_{a}\triangleright z^0-(T_{a}\triangleright z^1)\, e^{-q_0/\kappa}\frac{k_1}{\kappa} \,,\\
y^1-a^1&=& T_{a}\triangleright z^1\,,\\
w^0-a^0&=& T_{a}\triangleright z^0\,,\\
w^1-a^1e^{-q_0/\kappa}&=& ( T_{a}\triangleright z^1)\, e^{-q_0/\kappa} \,.
\eea
In order for the constraint equations to be covariant, the interaction parameter $z^\mu$ must transform classically:
\be
T_{a}\triangleright z^\mu=z^\mu-a^\mu\,.\label{translationz}
\ee
It turns out that the same is true for any number of incoming and outgoing particles.

In fact, it can be shown that the total relative locality action \eqref{eq:rel_loc_action_total} is invariant under such transformations \cite{AmelinoCamelia:2011nt}.
In particular, the equation defining the worldlines, Eq. \eqref{eq:rel_loc_worldline}, is covariant under translations generated by the total momentum, $T_{a}\triangleright \dot x^{\mu}_{I}=\dot x^{\mu}_{I}$, so the worldlines behave covariantly.

\subsubsection{Invariance under boosts}

Our next task is to identify the boost symmetry transformation that is compatible with the relative locality construction  in presence of one interaction vertex. This is the first time this issue is addressed.\footnote{Preliminary investigations \cite{Gubitosi:2013rna} overlooked the subtleties concerning the transformation properties of the momentum conservation constraint, Eq. \eqref{eq:rel_loc_conservation}.} In Section \ref{sub:backreaction} we have already shown that the conservation law of momenta is indeed invariant upon introducing a backreaction of momenta on the boost rapidity. Here we identify the way boosts act on spacetime coordinates $x^{\mu}_{I}$ and interaction coordinates $z^{\mu}$ that makes the constraint equation \eqref{eq:rel_loc_constraint_interaction}  transform covariantly. This is nontrivial due to the fact that the momentum conservation constraint function $\mathcal K_{\mu}$ does not transform covariantly (this is the technical realization of the nontrivial interplay between boosts and translations).

Since from the results on the conservation of momenta we know that the boost rapidity receives a backreaction from the momenta intervening in the interaction (see Subsection \ref{sub:backreaction}) one might be tempted to apply the same scheme to this problem.
Then the coordinates would transform as (compare to Eq. \eqref{eq:backreaction}):
\bea
x^{\mu}&\rightarrow& B^{\xi}\triangleright x^{\mu}\equiv x^{\mu}+\xi \{N,x^{\mu}\}\,,\\
y^{\mu}&\rightarrow&B^{\xi}\triangleright y^{\mu}\equiv y^{\mu}+\xi \{N,y^{\mu}\}\,,\\
w^{\mu}&\rightarrow&B^{\xi\triangleleft q}\triangleright w^{\mu}\equiv w^{\mu}+\xi e^{-\frac{q_0}{\kappa}} \{N,w^{\mu}\}\,,
\eea
where for all coordinates, $x^{\mu}$, $y^{\mu}$, $w^{\mu}$, the boost generator acts as on \eqref{eq:boosting_coordinates} (i.e. one can think of each boost as being represented in the phase space of the relevant spacetime coordinate and its conjugate momentum).
However, with this prescription there is no choice of transformation of the interaction coordinates $z^\mu$ which  leaves the constraint equations \eqref{eq:x0_constraint}-\eqref{eq:w1_constraint} invariant.
This can be understood once one realizes that the boost generators that are being used do not close the \kp algebra \eqref{eq:kp_algebra} with the  generators of translations given by the total momentum (which are the generators that transform the constraint equations covariantly).
And this observation is what guides us in identifying the boost symmetry transformations which leave the constraint equations invariant. 

Similarly to the case of translations, generated by the total momentum, in the case of boosts the symmetry generator is the  ``total boost", which was introduced in Subsection \ref{sub:backreaction}, see Eq. \eqref{eq:total_boost}. This is in fact the boost that closes the \kp algebra with the total momentum. Using such generator  the spacetime coordinates transform as follows:
\bea
x^{\mu}&\rightarrow& B^{\xi}\triangleright x^{\mu}\equiv x^{\mu}+\xi \{N_{[p]},x^{\mu}\} \label{eq:rel_loc_1vertex_xboosted}
 \,,\\
y^{\mu}&\rightarrow&B^{\xi}\triangleright y^{\mu}\equiv y^{\mu}+\xi \{N_{[q\oplus k]},y^{\mu}\}=y^{\mu}+\xi \{ N_{[q]}+ e^{-q_0/\kappa}N_{[k]},y^{\mu}\}\,,\\
w^{\mu}&\rightarrow&B^{\xi}\triangleright w^{\mu}\equiv w^{\mu}+\xi  \{N_{[q\oplus k]},w^{\mu}\}=w^{\mu}+\xi \{ N_{[q]}+ e^{-q_0/\kappa}N_{[k]},w^{\mu}\}  \label{eq:rel_loc_1vertex_wboosted}  \,.
\eea
Note that the representation of these boost generators is on the appropriate phase space coordinates. For example:
\be
N_{[q]} = q_1 y^0 + y^1 \left(\frac{\kappa}{2}\left(1-e^{-2 q_0/\kappa}\right)-\frac{ (q_1)^2}{2\kappa}\right)\,,
\ee
and analogously for $N_{[p]}$ and $N_{[k]}$.

Applying these ``total boost'' transformations to the constraint equations \eqref{eq:x0_constraint}-\eqref{eq:w1_constraint} one finds:
\bea
x^0-\xi e^{-2 p_0/\kappa}x^1&=& B^{\xi}\triangleright z^0\,, \\
x^1+\xi \left( \frac{p_1 x^1}{\kappa}-x^0  \right)&=& B^{\xi}\triangleright z^1 \,,\\
y^0-\xi e^{-2 q_0/\kappa}y^1 +\frac{\xi}{\kappa}N_{[k]} e^{- q_0/\kappa}  &=& B^{\xi}\triangleright \left(z^0-z^1 e^{-q_0/\kappa}\frac{k_1}{\kappa}\right) \,,\\
y^1 + \xi\left(\frac{q_1y^1}{\kappa} - y^0\right)&=& B^{\xi}\triangleright z^1\,,\\
w^0 - e^{-q_0/\kappa}\xi e^{-2k_0/\kappa}w^1&=& B^{\xi}\triangleright z^0\,,\\
w^1 + e^{-q_0/\kappa}\xi\left(\frac{k_1w^1}{\kappa} - w^0\right)&=& B^{\xi}\triangleright \left(e^{-q_0/\kappa}  z^1\right) \,,
\eea
where $N_{[k]}= k_1 w^0 + w^1 \left(\frac{\kappa}{2}\left(1-e^{-2 k_0/\kappa}\right)-\frac{ (k_1)^2}{2\kappa}\right)$. One can then verify that the constraint equations are covariant if the interaction parameter $z^\mu$ transforms as spacetime coordinates with the ``total boost"\footnote{Of course the ``total boost" must also be used in transforming the functions of momenta, such as in $B^{\xi}\triangleright \left(z^0-z^1 e^{-q_0/\kappa}\frac{k_1}{\kappa}\right)$.}:
\bea
B^{\xi}\triangleright z^0&=&z^0-\xi e^{-2 p_0/\kappa}z^1\,,\\
B^{\xi}\triangleright z^1&=&z^1+\xi \left( \frac{p_1 z^1}{\kappa}-z^0  \right)\,,
\eea
i.e. the boost acts on the interaction parameter in the same way as it would on the spacetime coordinates of a free particle with momentum $p_\mu$.

For a generic vertex with $n$ incoming and $m$ outgoing particles the same prescription applies. Coordinates of incoming particles transform with the total boost generator $N_{[p_{1}\oplus\dots\oplus p_{n}]}$, and coordinates of outgoing particles transform with the total boost $N_{[q_{1}\oplus\dots\oplus q_{m}]}$.

We can  show that also the worldlines of the particles are covariant with respect to a ``total boost'' transformation. We recall that the worldline associated to particle $I$ (either incoming or outgoing) is
\be
x^1_I=v(p_I)x^0_I\,,
\ee
 where we set the integration constant $\bar{x}^1_I=0$ for simplicity and  the velocity $v(p_I)$ is a function of momenta \emph{on shell}, see Eq. \eqref{eq:rel_loc_worldline1}.
 We want to verify that after a boost the worldline is still written in the same way:
 \be
(x^1_I)'=v(p_I')(x^0_I)'\,,
\ee
where prime denotes boosted variables.

 In the example we are studying, the incoming particle with momentum $p$  transforms as the free particle considered in the  previous Subsection, Eq. \eqref{eq:free_worldline_boost}. So the covariance of its worldline follows trivially from the invariance of a free particle worldline.
Concerning the two outgoing particles, we have to account for the fact that their coordinates and momenta transform according to the total momentum, so that their worldlines do not transform as the free particle worldline.
Considering for example the worldline of the particle with momentum $q$, this implies that the functional form of $B^\xi\triangleright y^0$ in terms of $y^0$, $y^1$ and $q_0$ is different from the functional form of $B^\xi\triangleright x^0$ in terms of $x^0$, $x^1$ and $p_0$. So  the boosted worldline of the particle with momentum $q$ has a different form compared to the boosted worldline of the particle with momentum $p$, Eq. \eqref{eq:free_worldline_boost}:
\be
y^1 + \xi\left(\frac{q_1y^1}{\kappa} - y^0\right) \stackrel{?}{=}v(q)\left(y^0-\xi e^{-2 q_0/\kappa}y^1 +\frac{\xi}{\kappa}N_{[k]} e^{- q_0/\kappa} \right) +\xi y^0 \{N_{[q\oplus k]},v(q)\} \,.\label{eq:interacting_worldline_boost}
\ee
Specifically, the difference resides in the additional term proportional to $N_{[k]}=k_1 w^0 + w^1 \left(\frac{\kappa}{2}\left(1-e^{-2 k_0/\kappa}\right)-\frac{ (k_1)^2}{2\kappa}\right)$ (because $\{N_{[q\oplus k]},v(q)\}=\{N_{[q]},v(q)\}$ is similar to what one has for the $p$ particle). However, one can easily check that $N_{[k]}=0$ when computed on the worldline of the particle with momentum $k$,
\be
w^1=v(k)w^0 \equiv \frac{e^{k_0/\kappa}\sqrt{e^{2 k_0/\kappa}+1-2e^{k_0/\kappa} \cosh(m/\kappa)}}{1-e^{k_0/\kappa}\cosh(m/\kappa)} w^0\,.
\ee
So for the worldline of the particle with momentum $q$ and coordinates $y$  relativistic invariance can be stated as the implication
\be
\left\{\begin{array}{lcl}	
y^1&=&v(q)\;y^0\\
w^1&=&v(k)\;w^0
\end{array}\right. \qquad \Leftrightarrow \qquad (y^1)'=v(q')(y^0)'\,,
\ee
where the equation on the right hand side is a compact version of Eq. \eqref{eq:interacting_worldline_boost}, now without the question mark.
Let us now investigate the invariance of the worldline of the particle with momentum $k$:
\be
w^1=v(k)w^0\,.
\ee
In this case neither $w^0$ nor $w^1$ transform as $x^0$ and $x^1$ respectively. In fact, applying a total boost to the two sides of the above equation gives
\be
w^1 + e^{-q_0/\kappa}\xi\left(\frac{k_1w^1}{\kappa} - w^0\right)\stackrel{?}{=}v(k)\left(w^0 - e^{-q_0/\kappa}\xi e^{-2k_0/\kappa}w^1\right)+\xi  \{N_{[q\oplus k]},v(k)\}\,.
\ee
This is however a much simpler case than the previous one. In fact, it is quite immediate to see that the difference with respect to how the worldline of the particle with momentum  $p$ is boosted resides in the extra factor  $e^{-q_0/\kappa}$ which multiplies all terms proportional to the rapidity. This overall factor is of course irrelevant and one can then state relativistic invariance in the usual way:
 \be
w^1=v(k)\;w^0\qquad \Leftrightarrow \qquad (w^1)'=v(k')(w^0)'\,.
\ee

Summarizing, we find that the worldlines of interacting particles, with one interaction vertex, are invariant under the action of the ``total boost", in a way that is informed about the fact that those worldlines are indeed interacting. This is because in order to show that the worldline of a given particle is invariant one might have to use information about the worldlines of the other particles intervening in the interaction, as was the case for the worldline of the particle with momentum $q$.

\subsection{Interacting model: multiple  vertices}

We are finally ready to tackle our last challenge: that of demonstrating the relativistic compatibility of the \kp relative locality model in presence of multiple causally connected interactions.  We have already discussed in Subsection \ref{sub:kPmultivertex} how the interplay between deformed boosts and translations becomes more involved in this case, because of the double role that translations have in defining the conservation rule of momenta and linking the different interaction vertices. In that Subsection we have also started to characterize the nontrivial relation between  distant and relatively boosted observers, which we  explore here in more depth.

To make the discussion explicit, we take again as example the process depicted in Figure \ref{fig:problem.diagram}. 
After constructing the phase space according to the relative locality prescription of Section \ref{sec:RL}, we see that in this process there is a finite worldline, that of particle with momentum $k$, whose endpoints are  at interaction vertex $1$ and interaction vertex $2$. So the additional complication with respect to the one-vertex case is that now relativistic invariance requires  that the transformation rules of the two endpoints are covariant and consistent with covariance of the worldline. This gives a consistency constraint on the  transformation rules of the interaction parameter $z^\mu$ associated to vertex $1$ and of the interaction parameter $\tilde z^\mu$, associated to vertex $2$. 

As suggested in \cite{AmelinoCamelia:2011nt}, we write the momentum conservation constraint functions $\mathcal K_\mu$ at the two vertices as differences of the total momentum before and after the interaction. Specifically, for  vertex $1$ we use $\mathcal K_\mu^{[1]}=p_\mu - (q \oplus k)_\mu$:
\bea
\mathcal K_0^{[1]}&=& p_0 - (q_0 + k_0)\,,\nonumber\\
\mathcal K_1^{[1]}&=&  p_1 - ( q_1 + e^{-q_0/\kappa}k_1)\,,
\eea
while for vertex $2$ we use $\mathcal K_\mu^{[2]}=(q\oplus k)_\mu - (q \oplus r \oplus s)_\mu$:
\bea
\mathcal K_0^{[2]}&=& (q_0+k_0) - (q_0+r_0 + s_0)\,,\nonumber\\
\mathcal K_1^{[2]}&=&  (q_1+e^{-q_0/\kappa} k_1 - ( q_1+e^{-q_0/\kappa} (r_1 + e^{-r_0/\kappa}s_1))\,.
\eea
Even though the contribution of the momentum $q$ in $K_\mu^{[2]}$ might seem redundant (and indeed it is at the level of the conservation rule of momenta), it is actually key in realizing translational invariance, as was demonstrated in \cite{AmelinoCamelia:2011nt} and reviewed in the following.

We denote by $x^\mu$, $y^\mu$, $w^\mu$, $u^\mu$, $v^\mu$ the spacetime coordinates dual to $p_\mu$, $q_\mu$, $k_\mu$, $r_\mu$, $s_\mu$, respectively (i.e. $\{x^\mu,p_\nu\}=\{y^\mu,q_\nu\}=\{w^\mu,k_\nu\}=\{u^\mu,r_\nu\}=\{v^\mu,s_\nu\}=\delta^\mu_\nu$). For the first vertex the boundary conditions at the endpoints of the worldlines are found via \eqref{eq:rel_loc_constraint_interaction} using $\mathcal K_\mu^{[1]}$:
\bea
x^0&=& z^0\,, \label{eq:x0_constraint_vertex1}\\
x^1&=& z^1 \,,\\
y^0&=& z^0-z^1 e^{-q_0/\kappa}\frac{k_1}{\kappa} \,,\\
y^1&=& z^1\,,\\
w^0&=& z^0\,, \label{eq:w0_constraint_vertex1}\\
w^1&=& z^1 e^{-q_0/\kappa}  \label{eq:w1_constraint_vertex1}\,,
\eea
while for the second vertex one uses $\mathcal K_\mu^{[2]}$ so that the constraint equations  read:
\bea
w^0&=&\tilde z^0\,, \label{eq:w0_constraint_vertex2}\\
w^1&=& e^{-q_0/\kappa} \tilde z^1 \,,\label{eq:w1_constraint_vertex2}\\
u^0&=& \tilde z^0- \tilde z^1 e^{-(r_0+q_0)/\kappa}\frac{s_1}{\kappa} \,,\\
u^1&=& \tilde z^1 e^{-q_0/\kappa}\,,\\
v^0&=& \tilde z^0\,,\\
v^1&=& \tilde z^1 e^{-(r_0+q_0)/\kappa}  \label{eq:v1_constraint}\,.
\eea
Note that for the sake of simplicity in the notation we are omitting the dependence on the value of the affine parameters $\lambda_{1}^{I}$  and $\lambda_{2}^{I}$ at the endpoints of the worldlines. Also,  the two sets of equations where $w^\mu$ appears, Eqs.\eqref{eq:w0_constraint_vertex1}-\eqref{eq:w1_constraint_vertex1} and Eqs. \eqref{eq:w0_constraint_vertex2}-\eqref{eq:w1_constraint_vertex2}, refer to the two different endpoints of the worldline of the particle with momentum $k$, so that in general $ z^0\neq \tilde z^0$ and $z^1\neq \tilde z^1$ (with a more complete notation one should  have written $w^\mu(\lambda_1^{w})$ and  $w^\mu(\lambda_2^{w})$ in the two cases respectively, making explicit reference to the values  the affine parameter $\lambda^{w}$ takes at the two endpoints of the worldline).

\subsubsection{Invariance under translations}

We briefly review the results first described in \cite{AmelinoCamelia:2011nt}. As for the one-vertex case, note that we are using a different convention for the Poisson brackets between spacetime coordinates and momenta, hence the specific formulas might look different, even though the results on translational invariance are equivalent.
Again, as for the one-vertex case, we use as translation generator the total momentum:
\bea
T_{a}\triangleright x^{\mu}&\equiv &x^{\mu}+a^{\nu}\{p_{\nu},x^{\mu}\}=x^{\mu}-a^{\mu}\,,\\
T_{a}\triangleright y^{\mu}&\equiv &y^{\mu}+a^{\nu}\{(q\oplus k)_{\nu},y^{\mu}\}=y^{\mu}-a^{\mu}+\delta^\mu_0a^1e^{-q_0/\kappa}\frac{k_1}{\kappa}\,,\\
T_{a}\triangleright w^{\mu}&\equiv &w^{\mu}+a^{\nu}\{(q\oplus k)_{\nu},w^{\mu}\}=w^{\mu}-a^{0}\delta^\mu_0-a^{1}e^{-q_0/\kappa}\delta^\mu_1\,,\\
T_{a}\triangleright u^{\mu}&\equiv &u^{\mu}+a^{\nu}\{(q\oplus r\oplus s)_{\nu},u^{\mu}\}=u^{\mu}-a^{0}\delta^\mu_0-a^{1}e^{-q_0/\kappa}\delta^\mu_1+\delta^\mu_0a^1e^{-(q_0+r_0)/\kappa}\frac{s_1}{\kappa}\,,\\
T_{a}\triangleright v^{\mu}&\equiv &v^{\mu}+a^{\nu}\{(q\oplus r\oplus s)_{\nu},v^{\mu}\}=v^{\mu}-a^{0}\delta^\mu_0-a^{1}e^{-(q_0+r_0)/\kappa}\delta^\mu_1\,.
\eea
Applying these transformations to the constraint equations \eqref{eq:x0_constraint_vertex1}-\eqref{eq:v1_constraint} one finds at vertex $1$:
\bea
x^0-a^0&=& T_{a}\triangleright  z^0\,, \\
x^1-a^1&=& T_{a}\triangleright z^1 \,,\\
y^0-a^0+a^1e^{-q_0/\kappa}\frac{k_1}{\kappa}&=& T_{a}\triangleright z^0-(T_{a}\triangleright z^1)\, e^{-q_0/\kappa}\frac{k_1}{\kappa} \,,\\
y^1-a^1&=& T_{a}\triangleright z^1\,,\\
w^0-a^0&=& T_{a}\triangleright z^0\,,\\
w^1-a^1e^{-q_0/\kappa}&=& (T_{a}\triangleright z^1)\,e^{-q_0/\kappa}  \,,
\eea
and at vertex $2$:
\bea
w^0-a^0&=& T_{a}\triangleright \tilde z^0\,, \\
w^1-a^{1}e^{-q_0/\kappa}&=&(T_{a}\triangleright  \tilde z^1)\,e^{-q_0/\kappa}  \,,\\
u^0-a^0+a^1e^{-(q_0+r_0)/\kappa}\frac{s_1}{\kappa}&=& T_{a}\triangleright \tilde z^0-(T_{a}\triangleright \tilde z^1) e^{-(q_0+r_0)/\kappa}\frac{s_1}{\kappa} \,,\\
u^1-a^1e^{-q_0/\kappa}&=& (T_{a}\triangleright \tilde z^1)\,e^{-q_0/\kappa} \,,\\
v^0-a^0&=& T_{a}\triangleright \tilde z^0\,,\\
v^1-a^{1}e^{-(q_0+r_0)/\kappa}&=& (T_{a}\triangleright \tilde z^1)\,e^{-(q_0+r_0)/\kappa}  \,.
\eea
So in order for the boundary conditions at the endpoints of the worldlines to be invariant both the interaction parameters $z^\mu$ and $\tilde z^\mu$ must transform classically:
\bea
T_{a}\triangleright z^\mu&=&z^\mu-a^\mu\,,\nonumber\\
T_{a}\triangleright \tilde z^\mu&=&\tilde z^\mu-a^\mu\,. \label{eq:rel_loc_ztildeztranslation}
\eea
It can also be shown that the total relative locality action \eqref{eq:rel_loc_action_total} is invariant under such transformations \cite{AmelinoCamelia:2011nt}, and in particular  the equation defining the worldlines, Eq. \eqref{eq:rel_loc_worldline}, is covariant.  Generalization to different combinations of vertices with any number of incoming and outgoing particles is straightforward (with the caveat already discussed in Subsection \ref{sub:kPmultivertex}), that only processes that  conserve total momentum are allowed).

In summary,  multiple-interaction systems are translational invariant if translations are generated by the total momentum, which does not  refer to just one given vertex, but accounts for all the particles that are causally connected.

\subsubsection{Invariance under boosts}

We are finally going to address the transformation properties under boosts. As was the case for the one-vertex scenario, this is the first time that the relativistic invariance under boost transformations of the  \kp relative locality framework with multiple interactions is discussed. 
Following a similar procedure as in the one-vertex case, we transform the spacetime coordinates using the ``total boost", which now accounts for all the causally connected particles and not just the ones directly entering a given vertex:
\bea
B^{\xi}\triangleright x^{\mu}&\equiv &x^{\mu}+\xi \{N_{[p]},x^{\mu}\} \label{eq:rel_loc_2vertex_xboosted}
 \,,\\
B^{\xi}\triangleright y^{\mu}&\equiv &y^{\mu}+\xi \{N_{[q\oplus k]},y^{\mu}\}=y^{\mu}+\xi \{ N_{[q]}+ e^{-q_0/\kappa}N_{[k]},y^{\mu}\}\,,\\
B^{\xi}\triangleright w^{\mu}&\equiv &w^{\mu}+\xi  \{N_{[q\oplus k]},w^{\mu}\}=w^{\mu}+\xi \{ N_{[q]}+ e^{-q_0/\kappa}N_{[k]},w^{\mu}\}   \label{eq:rel_loc_2vertex_wboosted} \,,\\
B^{\xi}\triangleright u^{\mu}&\equiv &u^{\mu}+\xi  \{N_{[q\oplus r\oplus s]},u^{\mu}\}=u^{\mu}+\xi \{ N_{[q]}+ e^{-q_0/\kappa}N_{[r]}+ e^{-(q_0+r_0)/\kappa}N_{[s]},u^{\mu}\}    \,,\\
B^{\xi}\triangleright v^{\mu}&\equiv &v^{\mu}+\xi  \{N_{[q\oplus r\oplus s]},v^{\mu}\}=v^{\mu}+\xi \{ N_{[q]}+ e^{-q_0/\kappa}N_{[r]}+ e^{-(q_0+r_0)/\kappa}N_{[s]},v^{\mu}\}    \,.
\eea
Applying these transformations to the constraint equations \eqref{eq:x0_constraint_vertex1}-\eqref{eq:v1_constraint} one finds at vertex $1$:
\bea
x^0-\xi e^{-2 p_0/\kappa}x^1&=& B^{\xi}\triangleright z^0\,, \label{eq:x0_boosted_2vert}\\
x^1+\xi \left( \frac{p_1 x^1}{\kappa}-x^0  \right)&=& B^{\xi}\triangleright z^1 \,,\\
y^0-\xi e^{-2 q_0/\kappa}y^1 +\frac{\xi}{\kappa}N_{[k]} e^{- q_0/\kappa}  &=& B^{\xi}\triangleright \left(z^0-z^1 e^{-q_0/\kappa}\frac{k_1}{\kappa}\right) \,,\\
y^1 + \xi\left(\frac{q_1y^1}{\kappa} - y^0\right)&=& B^{\xi}\triangleright z^1\,,\\
w^0 - e^{-q_0/\kappa}\xi e^{-2k_0/\kappa}w^1&=& B^{\xi}\triangleright z^0\,,\\
w^1 + e^{-q_0/\kappa}\xi\left(\frac{k_1w^1}{\kappa} - w^0\right)&=& B^{\xi}\triangleright \left(z^1 e^{-q_0/\kappa}  \right) \,,
\eea
and at vertex $2$: 
\bea
w^0 - e^{-q_0/\kappa}\xi e^{-2k_0/\kappa}w^1&=& B^{\xi}\triangleright \tilde z^0\,, \\
w^1 + e^{-q_0/\kappa}\xi\left(\frac{k_1w^1}{\kappa} - w^0\right)&=&B^{\xi}\triangleright ( \tilde z^1 e^{-q_0/\kappa}  ) \,,\label{eq:w1_boosted_2vert}\\
u^0-\xi e^{- q_0/\kappa}e^{- 2r_0/\kappa}u^1 +\frac{\xi}{\kappa}N_{[s]} e^{- (q_0+r_0)/\kappa}  &=& B^{\xi}\triangleright ( \tilde z^0- \tilde z^1 e^{-(q_0+r_0)/\kappa}\frac{s_1}{\kappa}) \,,\\
u^1 + \xi e^{- q_0/\kappa} \left(\frac{r_1u^1}{\kappa} - r^0\right)&=& B^{\xi}\triangleright (\tilde z^1 e^{-q_0/\kappa} )\,,\\
v^0 - \xi e^{- (q_0+r_0)/\kappa}e^{-2s_0/\kappa}v^1&=& B^{\xi}\triangleright  \tilde z^0\,,\\
v^1 + \xi   e^{- (q_0+r_0)/\kappa}\left(\frac{s_1v^1}{\kappa} - v^0\right)&=& B^{\xi}\triangleright (\tilde z^1 e^{-(q_0+r_0)/\kappa}  ) \,.\label{eq:v1_boosted_2vert}
\eea

One can then verify that the constraint equations are invariant if the interaction parameters $z^\mu$ and $\tilde z^\mu$ transform as spacetime coordinates with the ``total boost":
\bea
B^{\xi}\triangleright z^0&=&z^0-\xi e^{-2 p_0/\kappa}z^1\,, \label{eq:rel_loc_boostedz0}\\
B^{\xi}\triangleright z^1&=&z^1+\xi \left( \frac{p_1 z^1}{\kappa}-z^0  \right)\,,\label{eq:rel_loc_boostedz1}\\
B^{\xi}\triangleright \tilde z^0&=&\tilde z^0-\xi e^{-2 p_0/\kappa}\tilde z^1\,,\label{eq:rel_loc_boostedz0tilde}\\
B^{\xi}\triangleright \tilde z^1&=&\tilde z^1+\xi \left( \frac{p_1 \tilde z^1}{\kappa}-\tilde z^0  \right)\,.\label{eq:rel_loc_boostedz1tilde}
\eea

Note again that  the ``total boost" rule also works on the right-hand side of Eqs. \eqref{eq:x0_boosted_2vert}-\eqref{eq:v1_boosted_2vert}. For example, the right-hand side of Eq. \eqref{eq:w1_boosted_2vert} gives:
\bea
B^{\xi}\triangleright (e^{-q_0/\kappa}   \tilde z^1) &=&e^{-q_0/\kappa}   \tilde z^1+\xi \{N_{[q\oplus k]},e^{-q_0/\kappa}   \tilde z^1\}=e^{-q_0/\kappa}   \tilde z^1+\xi e^{-q_0/\kappa} \{N_{[q\oplus k]},  \tilde z^1\}+\xi \{N_{[q\oplus k]},e^{-q_0/\kappa}\}   \tilde z^1\nonumber\\
&=&e^{-q_0/\kappa}   \tilde z^1+\xi e^{-q_0/\kappa}\left( \frac{(q\oplus k)_1 \tilde z^1 }{\kappa}- \tilde z^0\right)+\xi \{N_{[q]},e^{-q_0/\kappa}\}   \tilde z^1\nonumber \\
&=& e^{-q_0/\kappa}   \tilde z^1+\xi e^{-q_0/\kappa}\left( \frac{e^{-q_0/\kappa} k_1 \tilde z^1 }{\kappa}- \tilde z^0\right)\,.
\eea

Now that we have the transformation rule of the coordinates that guarantees the covariance of the boundary equations at the endpoints of the worldlines, we are only left with the task of verifying that these transformations also leave the worldlines themselves invariant.

The results from the one-vertex case in the previous Subsection  showed that the worldlines of the particles with momenta $p$, $q$ and $k$ are invariant. Since the action of boosts on spacetime coordinates and momenta of these particles does not change when  gluing vertex $2$ to the worldline of the particle with momentum $k$ (compare Eqs.\eqref{eq:rel_loc_1vertex_xboosted}-\eqref{eq:rel_loc_1vertex_wboosted} and Eqs. \eqref{eq:rel_loc_2vertex_xboosted}-\eqref{eq:rel_loc_2vertex_wboosted}), we can import the results on the invariance of the  worldlines to this case. We are thus left only with the task of demonstrating the invariance of the worldlines of particles with momenta $r$ and $s$: 
\bea
u^1&=&v(r)u^0\,,\\
v^1&=&v(s)v^0\,.
\eea
The demonstration  follows a very similar argument to the one used for particles with momenta $q$ and $k$. Concerning the particle with momentum $r$, acting with the ``total boost" $N_{[q\oplus r\oplus s]}$ on the worldline of this particle one finds:
\be
u^1 + \xi e^{- q_0/\kappa} \left(\frac{r_1u^1}{\kappa} - r^0\right)\stackrel{?}{=}v(r)(u^0-\xi e^{- q_0/\kappa}e^{- 2r_0/\kappa}u^1 +\frac{\xi}{\kappa}N_{[s]} e^{- (q_0+r_0)/\kappa} )+\xi u^0 \{N_{[q\oplus r\oplus s]},v(r)\}\,,
\ee
where $\{N_{[q\oplus r\oplus s]},v(r)\}=\{N_{[q]}+e^{-q_0/\kappa}N_{[r]}+e^{-(q_0+r_0)/\kappa}N_{[s]},v(r)\}=e^{-q_0/\kappa}\{N_{[r]},v(r)\}$. The term $N_{[s]}$ vanishes when computed on the worldline of the particle with momentum $s$,   $N_{[s]}|_{v^1=v(s)v^0}=0$. All the other terms proportional to $\xi$ cancel when computed on the worldline of the particle with momentum $r$, so that the invariance of the worldline of this particle reads:
\be
\left\{\begin{array}{lcl}	
u^1&=&v(r)\;u^0\\
v^1&=&v(s)\;v^0
\end{array}\right. \Leftrightarrow  (u^1)'=v(r')(u^0)'\,.
\ee
Invariance of the worldline of the particle with momentum $s$ follows analogously. So again we find that  the invariance under boost transformations of the worldlines of interacting particles cannot be demonstrated independently for each particle, but requires information from the whole causally connected system.

\subsubsection{Second interlude:  back to the composition of deformed symmetry transformations}

Having constructed  the full phase space of the \kp  interacting model, and having understood how relativistic symmetries are realized in this framework, we are now fully equipped to give a more precise characterization of the interplay between translation and boost transformations which was discussed in the interlude of Subsection \ref{sub:kPmultivertex}, since we can now use interaction events to  define different observers.

Let us again make reference to the interaction processes depicted in Figure \ref{fig:problem.diagram} and define four observers, $A, B, C, D$.
\begin{itemize}

\item Observer A  is local to interaction $1$, so for this observer $z_{A}^{\mu}=0$, and as a consequence of Eqs. \eqref{eq:x0_constraint_vertex1}-\eqref{eq:w1_constraint_vertex1}, $x^{\mu}_{A}(\lambda_{1}^{p})=y^{\mu}_{A}(\lambda_{1}^{q})=w^{\mu}_{A}(\lambda_{1}^{k})=0$ (the index A means that this is the value of the quantity as seen by observer A).
For this observer of course $\tilde z_{A}^{\mu}\neq 0$, since the second interaction happens at some distant point along the worldline of the  particle with momentum $k$:  $w^{\mu}_{A}(\lambda_{2}^{k})=T_{a}\triangleright w^{\mu}_{A}(\lambda_{1}^{k})\neq 0$.
\item
Observer $B\equiv T_{\tilde z_{A}}\triangleright A$ (translated with respect to A with a translation parameter $a=\tilde z_{A}$) sees the interaction $2$ as local: $\tilde z_{B}^{\mu}=0 \Rightarrow w^{\mu}_{A}(\lambda_{2}^{k})=u^{\mu}_{A}(\lambda_{2}^{r})=v^{\mu}_{A}(\lambda_{2}^{s})=0$ (see Eqs. \eqref{eq:rel_loc_ztildeztranslation} and \eqref{eq:w0_constraint_vertex2}-\eqref{eq:v1_constraint}). This observer of course sees interaction $1$ as nonlocal, $z_{B}\neq 0$.
\item Observer $C\equiv B^{\xi}\triangleright A$ is local to the interaction vertex $1$, but boosted with respect to $A$. Since $z_{A}=0$ then also $z_{C} = B^{\xi}\triangleright z_{A} = 0$ (see Eqs. \eqref{eq:rel_loc_boostedz0}-\eqref{eq:rel_loc_boostedz1}). This observer also sees the interaction $2$ as nonlocal, but with a different amount of nonlocality as compared to the one seen by A: 
\be
\tilde z_{C}=B^{\xi}\triangleright \tilde z_{A}\Rightarrow\left\{ \begin{array}{lcl}
\tilde z_{C}^{0}&=& \tilde z_{A}^{0}-\xi e^{-2 p_{0}/\kappa}\tilde z^{1}_{A}\\
\tilde z_{C}^{1}&=& \tilde z_{A}^{1}+\xi \left( \frac{p_{1} \tilde z^{1}_{A}}{\kappa}-\tilde z^{0}_{A}\right)
\end{array}  \,, \right.\label{eq:observerC}
\ee
(see Eqs. \eqref{eq:rel_loc_boostedz0tilde}-\eqref{eq:rel_loc_boostedz1tilde}).
\item The fourth observer, $D$, is defined as the one that is translated with respect to $C$ and sees the interaction $2$ as local. To identify this observer we need to determine the value of the translation parameter $a$ such that $\tilde z_{D}\equiv T_{a}\triangleright \tilde z_{C}=0$. Applying the translation transformation \eqref{eq:rel_loc_ztildeztranslation} to Eq. \eqref{eq:observerC} one finds that the translation parameter $a$ must satisfy the following conditions:
\bea
a^{0}&=& \tilde z_{A}^{0}-\xi e^{-2 p_{0}/\kappa}\tilde z^{1}_{A} = B^{\xi} \triangleright \tilde z_{A}^{0}\,,\\
a^{1}&=& \tilde z_{A}^{1}+\xi \left( \frac{p_{1} \tilde z^{1}_{A}}{\kappa}-\tilde z^{0}_{A}\right)=B^{\xi} \triangleright \tilde z_{A}^{1}\,.
\eea
And actually this same observer $D$ can be defined by boosting observer $B$: $D=B^{\xi}\triangleright B$ (one can indeed check that the amount of nonlocality seen at interaction $1$ by $B^{\xi}\triangleright B$ is the same  seen  by $D$ as defined originally).
So the observer $D$ can be reached from $A$ in two ways:
\bea
D&=& T_{B^{\xi} \triangleright \tilde z_{A}}\triangleright\left( B^{\xi}\triangleright A \right)\,,\\
D&=&  B^{\xi}\triangleright\left(T_{\tilde z_{A}} \triangleright A \right)\,.\\
\eea
\end{itemize}
The momentum dependence of the translation parameter defining observer $D$ and the two possible ways of reaching observer $D$ from $A$ are a concrete physical realization  of the results we had found in the interlude of Subsection \ref{sub:kPmultivertex} when studying the commutation between a boost and a translation, Eqs. \eqref{eq:TBcommutation} and \eqref{eq:momentumdependenttranslation},   showing that \kp symmetry transformations can only be fully defined in the whole phase space. Moreover we have uncovered the fact that the momentum dependent translation transformation  $T_{B^{\xi} \triangleright \tilde z_{A}}$ does not depend on the momentum of the worldline at whose endpoint  observers $A$ and $D$ lie, but on the total momentum of the system, $p=q\oplus k$.

\section{Conclusions and outlook}

This paper contributes to  the understanding of how deformed relativistic symmetries can be realized in models with nontrivial momentum space geometry, whose relevance for quantum gravity research is increasing  from both the theoretical and phenomenological perspectives.

For our investigations we relied on the much studied  $\kappa$-Poincar\'e model, both in its momentum space realization and in the complete phase space picture provided by the relative locality framework. We demonstrated the full relativistic compatibility of the model not just in the free-particle case, which was already well understood,  but also  in presence of several causally connected interactions. We focussed specifically on understanding how Lorentz invariance is achieved in systems of interacting particles, where there is a nontrivial interplay with the translational invariance,  affecting both the composition rule of momenta in interactions and  the connection between different interaction vertices.

 We showed that the way a boosted observer sees the phase space of a  particle that  belongs to a system of   several interactions depends on the properties of all particles that are causally connected to the particle itself, even if they do not directly interact with it. This can be formalized by stating that the action of a boost transformation on a system of particles is given by the ``total boost'' generator, which is a nontrivial sum of the boost generators acting on single particles (the form of this sum is induced by the coproduct of the boost generator in the \kp Hopf algebra). In the case of one single interaction vertex, this was previously understood as a ``backreaction'' of the momenta of the particles onto the boost rapidity. We showed here that while this latter interpretation is fully equivalent to the ``total boost'' action when focussing on the momentum space picture,  its limitations emerge when introducing spacetime and the particles worldlines.
 
 As a byproduct of our analysis, we were able to define a network of  causally connected events compatible with the relativistic symmetries of the model. This allowed us to  define different kinds of observers, each local to different interaction events and with a different amount of boost. We showed that the transformation relating distant and relatively boosted observers is actually a phase space transformation (as opposed to a purely spacetime or momentum space transformation) and we computed the amount of non-locality that each observer sees in  interactions belonging to a causally connected chain.

 We note that in order to construct a boost invariant relative locality phase space with interactions is was crucial to use the properties of boost generators that are provided by the \kp Hopf algebra structure, which determined the way  single-particle boost generators enter into the sum producing the ``total boost''. We wonder whether more general momentum spaces, such as the ones studied in \cite{Amelino-Camelia:2013sba, Arzano:2014jua} do have a rich enough structure to guarantee Lorentz invariance of interactions. 
 
 Somewhat related to this, now that a fully relativistic picture of the phase space of interacting particles with curved momentum sector and flat spacetime sector is available, we hope in the future to build a similar framework for models, still based on Hopf-algebraic symmetries, where also spacetime is curved. Such models are at a much earlier stage of development, since we are still in the process of  understanding  the free-particle phase space structure \cite{Barcaroli:2015eqe, Ballesteros:2017kxj, Ballesteros:2017pdw}. We feel that our contribution to understanding the way in which the structures of the underlying Hopf algebra enter into the physical realization of the relativistic symmetries might be of guidance in this endeavour.

Finally, our findings raise several questions concerning the observational implications of the \kp model, which will be further studied in future work. For example, one might wonder how to deal with the ``total boost'' transformation law in realistic situations where one does not know the full  chain of interactions to which a particle might be causally connected. Maybe one could infer what  interactions have generated a particle of astrophysical origin by looking at  how the particle is seen by different relatively boosted observers? And what are  physically relevant scenarios where the  consequences of the nontrivial composition of translations and boost transformations would be observable?

\section*{Acknowledgements}
GG acknowledges support from the Junta de Castilla y Le\'on (Spain) under grant BU229P18.

\bibliographystyle{apsrev4-1}   
\bibliography{sjors1}

\end{document}